\documentclass[fleqn,usenatbib,useAMS]{mnras}

\usepackage{graphicx}	
\usepackage{amsmath}	
\usepackage{amssymb}	
\usepackage{multicol}        
\usepackage{bm}		
\usepackage{pdflscape}	
\usepackage{soul}
\usepackage{enumitem}
\usepackage[dvipsnames]{xcolor}

\usepackage[T1]{fontenc}
\usepackage{ae,aecompl}

\usepackage{newtxtext,newtxmath}

\newcommand{\Mpc}{\mathrm{Mpc}}
\newcommand{\Myr}{\mathrm{Myr}}
\newcommand{\Gyr}{\mathrm{Gyr}}
\newcommand{\kpc}{\mathrm{kpc}}
\newcommand{\pc}{\mathrm{pc}}
\newcommand{\cmcube}{\mathrm{cm}^{-3}}
\newcommand{\protonmass}{m_\mathrm{p}}
\newcommand{\mppercmcube}{\protonmass \, \cmcube}

\newcommand{\Msol}{\textup{M}_\mathrm{\sun}}

\newcommand{\xMsol}[2]{\ensuremath{{#1}\times 10^{#2} \,\Msol}}
\newcommand{\xScientific}[2]{\ensuremath{{#1} \times 10^{#2}}}
\newcommand{\Msolyr}{\textup{M}_\mathrm{\sun} \, \text{yr}^{-1}}

\newcommand{\hi}{H{\textsc{i}}}

\newcommand{\Mvir}{M_{200}}

\newcommand{\Tvir}{T_{200}}
\newcommand{\Mgas}{M_{\text{gas}}}
\newcommand{\Mstar}{M_{\star}}
\newcommand{\rvir}{r_{200}}
\newcommand{\magv}{\mathcal{M}_V}
\newcommand{\rhalflight}{r_{1/2}}
\newcommand{\epsilonff}{\epsilon_{\text{ff}}}
\newcommand{\tff}{t_{\text{ff}}}
\newcommand{\Tigm}{T_{\text{IGM}}}

\newcommand{\MR}[1]{{\color{black}{#1}}}

\title[\MR{Gas-rich to star-forming dwarfs}]
{EDGE: From quiescent to gas-rich to star-forming low-mass \\ dwarf galaxies}

\author[M. P. Rey et al.]
{Martin P. Rey$^{1,\,2}$\thanks{Contact e-mail: \href{martin.rey@astro.lu.se}{martin.rey@astro.lu.se}}, Andrew Pontzen$^2$, Oscar Agertz$^1$, Matthew D. A. Orkney$^3$, \newauthor Justin I. Read$^3$, Joakim Rosdahl$^4$
\vspace{0.8mm}
\\
$^{1}$Lund Observatory, Department of Astronomy and Theoretical Physics, Lund University, Box 43, SE-221 00, Lund, Sweden \\
$^{2}$Department of Physics and Astronomy, University College London, London WC1E 6BT, UK \\
$^{3}$Department of Physics, University of Surrey, Guildford GU2 7XH, UK
\\
$^{4}$Univ Lyon 1, Ens de Lyon, CNRS, Centre de Recherche Astrophysique de Lyon, UMR5574, F-69230, Saint-Genis-Laval, France
}

\date{Submitted to MNRAS}

\pubyear{2020}

\begin{document}
\label{firstpage}
\pagerange{\pageref{firstpage}--\pageref{lastpage}}
\maketitle

\begin{abstract}
  We study how star formation is regulated in low-mass field dwarf galaxies ($10^5 \leq M_{\star} \leq 10^6 \, \textup{M}_\mathrm{\sun}$), using cosmological high-resolution ($3 \, \mathrm{pc}$) hydrodynamical simulations. Cosmic reionization quenches star formation in all our simulated dwarfs, but three galaxies with final dynamical masses of $3 \times 10^{9} \,\textup{M}_\mathrm{\sun}$ are subsequently able to replenish their interstellar medium by slowly accreting gas. Two of these galaxies reignite and sustain star formation until the present day at an average rate of $10^{-5} \, \textup{M}_\mathrm{\sun} \, \text{yr}^{-1}$, highly reminiscent of observed low-mass star-forming dwarf irregulars such as Leo~T. The resumption of star formation is delayed by several billion years due to residual feedback from stellar winds and Type Ia supernovae; even at $z=0$, the third galaxy remains in a temporary equilibrium with a large gas content but without any ongoing star formation. Using the `genetic modification' approach, we create an alternative mass growth history for this gas-rich quiescent dwarf and show how a small $(0.2\,\mathrm{dex})$ increase in dynamical mass can overcome residual stellar feedback, reigniting star formation. The interaction between feedback and mass build-up produces a diversity in the stellar ages and gas content of low-mass dwarfs, which will be probed by combining next-generation H{\textsc{i}} and imaging surveys.
\end{abstract}




\section{Introduction}

Cosmic reionization plays a key role in shaping the properties of faint dwarf galaxies. By $z=6$, the Universe has been reionized by ultraviolet (UV) photons emitted from the entire population of galaxies and quasars, providing a near-uniform heating and ionizing background of radiation across the Universe (see \citealt{McQuinn2016_review} for a review). Heating from the UV background raises the Jeans mass of the intergalactic medium (IGM), hence preventing the condensation of gas onto haloes with dynamical masses below a few $10^9 \, \Msol$ (\citealt{Efstathiou1992, Shapiro1994, Noh2014}). Dwarf galaxies are unable to accrete gas from the IGM below this dynamical mass threshold, leading to a suppression or termination of star formation (e.g. \citealt{Gnedin2000, Hoeft2006, Okamoto2008, Noh2014}). Such reionization-driven regulation of star formation has been crucial for explaining the faintest end of the luminosity function (\citealt{Bullock2000, Benson2002, Somerville2002}) and the observed ages of stars within ultrafaint dwarf galaxies (e.g. \citealt{Okamoto2012, Brown2014, Weisz2014}, see \citealt{Simon2019} for a review).
\begin{table*}
  \centering
     \begin{tabular}{l c c c c c c c}
     \hline
     Simulation name & $\Mvir$ ($\Msol$)& $\Mstar$ ($\Msol$) & $\Mgas(<1 \, \kpc)$ ($\Msol$) & $\magv$ & $\rhalflight$ ($\kpc$) & \textbf{SF at $z=0$}\\
     \hline

     \textcolor{MidnightBlue}{Halo 600}: rejuvenated, fast growth& $\xScientific{3.3}{9}$ & $\xScientific{5.1}{5}$ & $\xScientific{1.8}{5}$ & -8.8 & 0.17 & \textbf{Yes}\\

     \quad $\hookrightarrow$ GM: delayed mergers & $\xScientific{3.2}{9}$ & $\xScientific{3.9}{5}$ & $\xScientific{7.2}{5}$ & -8.2 & 0.20 & \textbf{Yes} \\
     
     \quad $\hookrightarrow$ Earlier reionization & $\xScientific{3.4}{9}$ & $\xScientific{1.6}{5}$ & $\xScientific{5.}{5}$ & -7.7 & 0.12 & \textbf{Yes} \\

     \quad $\hookrightarrow$ Weak feedback & $\xScientific{3.4}{9}$ & $\xScientific{4.2}{6}$ & $\xScientific{1.7}{7}$ & -13.0 & 0.26 & \textbf{Yes} \\

     \hline

     \textcolor{RedViolet}{Halo 605}: rejuvenated, steady growth& $\xScientific{3.2}{9}$ & $\xScientific{1.7}{6}$ & $\xScientific{9.2}{5}$ & -9.7 & 0.21 &\textbf{Yes}\\

     \hline

     \textcolor{RawSienna}{Halo 624}: quenched, gas-rich & $\xScientific{2.5}{9}$ & $\xScientific{5.9}{5}$ & $\xScientific{1.0}{6}$ & -8.9 & 0.26 & No\\

     \quad $\hookrightarrow$ GM: higher final mass & $\xScientific{3.7}{9}$ & $\xScientific{1.5}{6}$ & $\xScientific{5.3}{5}$ & -9.6 & 0.23 & \textbf{Yes} \\

     \quad $\hookrightarrow$ No AGB & $\xScientific{2.5}{9}$ & $\xScientific{6.0}{5}$ & $\xScientific{9.6}{5}$ & -9.0 & 0.26 & \textbf{Yes} \\

     \quad $\hookrightarrow$ No AGB, no SNIa  & $\xScientific{2.5}{9}$ & $\xScientific{6.0}{5}$ & $\xScientific{1.0}{6}$ & -9.0 & 0.24 & \textbf{Yes} \\

     \hline

     Halo 1459: quenched & $\xScientific{1.4}{9}$ & $\xScientific{5}{5}$ & $\xScientific{2}{3}$ & -8.2 & 0.13 & No\\

     \hline
     \end{tabular}
    \vspace*{4mm}

    \caption{Simulations presented in this work and their dynamical halo masses, stellar masses, gas masses within the inner $1\, \kpc$, total absolute $V$-band magnitudes, and half-light radius at $z=0$ (second, third, fourth, fifth and sixth column respectively). The last column indicates whether the galaxy has formed a stellar population in the $500\, \Myr$ before the end of the simulation. We present four independent galaxies separated by horizontal lines evolved with our fiducial galaxy formation model (Section~\ref{sec:setup:ramses}). We further explore the sensitivity of our results to modifying the mass accretion history of galaxies using the genetic modification approach (labelled GM -- see Section~\ref{sec:results:gms}) and galaxy formation physics (see Section~\ref{sec:results:residualFB} and~\ref{sec:results:physics}).}
   \label{table:runs}
\end{table*}

The discovery of multiple faint low-mass dwarf galaxies, which are currently forming stars with rates as low as $10^{-5} \, \Msolyr$, is a unique challenge to the above scenario. Thanks to its early discovery and close proximity, Leo~T is the best-studied example of such galaxies. Its low stellar mass, typical of ultrafaint dwarfs ($\Mstar \sim 10^5 \, \Msol$, $\magv \sim -7$) is contrasted by a young stellar population born in the last $200 \, \Myr$ (\citealt{Irwin2007}). Deep colour--magnitude measurements further reveal that Leo~T has formed stars at a constant rate for, at least, the past $6 \, \Gyr$ (\citealt{deJong2008, Clementini2012, Weisz2012}). Radio observations show an extended \hi \ reservoir that dominates over the mass of the stellar body, and with a cold and warm phase as expected from an interstellar medium (\citealt{RyanWeber2008, Adams2018}). The lack of coherent \hi \ rotation makes it challenging to determine Leo~T's dynamical mass (\citealt{Adams2018, Patra2018}), and stellar kinematics remain consistent with dynamical masses up to $\xScientific{3}{10} \, \Msol$ (e.g. \citealt{Errani2018, Forbes2018}). However, the low levels of star formation and stellar mass of Leo~T both suggest a dynamical mass smaller than  $\xScientific{3}{9} \, \Msol$ using abundance-matching arguments (\citealt{Read2017, Read2019}). It remains an open question how such a low-mass object would sustain star formation after reionization (e.g. \citealt{Ricotti2009, Wright2019}).

It is unclear whether Leo~T's characteristics are unusual amongst the dwarf galaxy population; rarity may indicate a physically unusual formation scenario, or could simply reflect the challenge of finding such low-surface-brightness dwarfs in the field. One way to distinguish these possibilities is to search for similar objects using their relatively large gas content. Within wide \hi \ surveys of the Local Volume, catalogues of \hi \ selected clouds can be matched or followed up with deep photometric imaging. When applying this approach with the GALFA-\hi \ (\citealt{Peek2011}) and ALFALFA (\citealt{Haynes2011}) surveys, \citet{DeFelippis2019} and \citet{Janesh2019}, respectively, uncovered zero and five (\citealt{Janesh2019}) gas-rich ultrafaint dwarf candidates. Furthermore, a similar \hi-selection has repeatedly exposed new star-forming dwarf galaxies with slightly higher masses ($\Mstar \leq 10^6 \, \Msol$; \citealt{McQuinn2015LeoP, Brunker2019, Hargis2020, McQuinn2020}). Our knowledge of nearby gas-rich dwarf galaxies is therefore unlikely to be complete, but a clearer census of the population could be revealed in the near future, given forthcoming deep optical imaging (notably \MR{the Rubin Observatory Legacy Survey of Space and Time, hereafter LSST}\footnote{https://www.lsst.org/.}). There is a strong motivation to perform such cross-matching; determining the scale at which low-mass dwarf galaxies can sustain star formation is key to interpret observations aiming at constraining the nature of dark matter (e.g. \citealt{Pontzen2012, DiCintio2014}) and cosmic reionization (e.g. \citealt{Ricotti2005, Tollerud2018}). 

In preparation for these forthcoming observations, this paper studies the ability of low-mass dark matter haloes to play host to Leo T-like galaxies. As with all studies of dwarf galaxies, a key factor is how the cosmological mass accretion interacts with reionization and astrophysical processes within the interstellar medium (e.g. \citealt{Maccio2017,  Revaz2018, Munshi2019, Wheeler2019, Agertz2020}). To tackle this, we use a suite of high-resolution ($3 \, \pc$) cosmological simulations from the Engineering Dwarfs at Galaxy Formation's Edge (EDGE) project (\citealt{Agertz2020}). In addition to high resolution and detailed astrophysical subgrid mechanisms, this suite includes genetically modified initial conditions \citep{Roth2016, Rey2018}, which allows us to construct several alternative versions of each galaxy. Each version modifies a specified aspect of the mass accretion history to establish a causal link between a galaxy's history and its observables (\citealt{Pontzen2017, Rey2019b}). In comparison with previous studies of the interaction between reionization and star formation (e.g. \citealt{Ricotti2005, Okamoto2008, BenitezLLambay2015, Sawala2016, Fitts2017, Ledinauskas2018}), we target dark matter haloes with lower dynamical masses ($\xScientific{1-3}{9} \, \Msol$), which are more likely to host observed ultrafaint dwarfs such as Leo~T (\citealt{Jeon2017, Jethwa2018, Read2019}). 

We describe how we construct our suite of simulated dwarf galaxies in Section~\ref{sec:setup}, before showing that our most massive galaxies exhibit a reignition of star formation at late times and are forming new stars until $z=0$ (Section~\ref{sec:results}). We demonstrate that the reignition of star formation is explained by an isothermal accretion of gas, until self-shielding is reached in the halo centre (Section~\ref{sec:results:cooling}). We also show that residual feedback from stellar winds and Type Ia supernovae (SNe Ia) has a significant role to play in regulating the neutral gas content within such low-mass objects (Section~\ref{sec:results:residualFB}). We discuss the sensitivity of our result to both the specific mass build-up of our galaxies and uncertainties in the subgrid models (Sections~\ref{sec:results:gms} and~\ref{sec:results:physics}), before concluding in Section~\ref{sec:conclusion}.

\section{Numerical setup} \label{sec:setup}

We start by describing our suite of simulated galaxies. We present a series of high-resolution, zoomed cosmological galaxy formation simulations (\citealt{Agertz2020}), complemented by the genetic modification approach (e.g. \citealt{Rey2019b}). We describe how we select and construct the cosmological initial conditions (Section~\ref{sec:setup:ics}) and the galaxy formation model with which they are evolved to $z=0$ (Section~\ref{sec:setup:ramses}). Properties of the simulated galaxies presented in this paper are presented in Table~\ref{table:runs}.

\subsection{Initial conditions}   \label{sec:setup:ics}

We construct zoom initial conditions for four dark matter haloes within a narrow window in present-day halo mass, ranging from $\Mvir = \xScientific{1.5}{9}$ to $\xMsol{3.5}{9}$, where $\Mvir$ defines the mass enclosed within a sphere of radius, $\rvir$, encompassing $200$ times the critical density of the Universe. We use the same procedure as in \citet{Agertz2020} -- briefly, we first simulate a dark-matter-only cosmological volume ($50 \, \Mpc$ at $512^3$ resolution), before re-simulating the largest void in the simulation with a zoomed dark-matter-only simulation (resolution equivalent to $2048^3$). We identify haloes within the void using the \textsc{hop} halo finder (as in \citealt{Eisenstein1998}) and select four haloes for our main suite as isolated centrals, with no more massive neighbours within $5 \, \rvir$. We track these haloes to the initial conditions again, and re-simulate with hydrodynamics and galaxy formation physics at our final, zoomed resolution equivalent to $16384^3$ (i.e. $m_{\text{DM}} = 960 \, \Msol$ and \MR{and $m_{\star} = 300 \, \Msol$, where $m_{\text{DM}}$ and $m_{\star}$ are the dark matter and stellar particle masses, respectively}).

\begin{figure}
  \centering
    \includegraphics[width=\columnwidth]{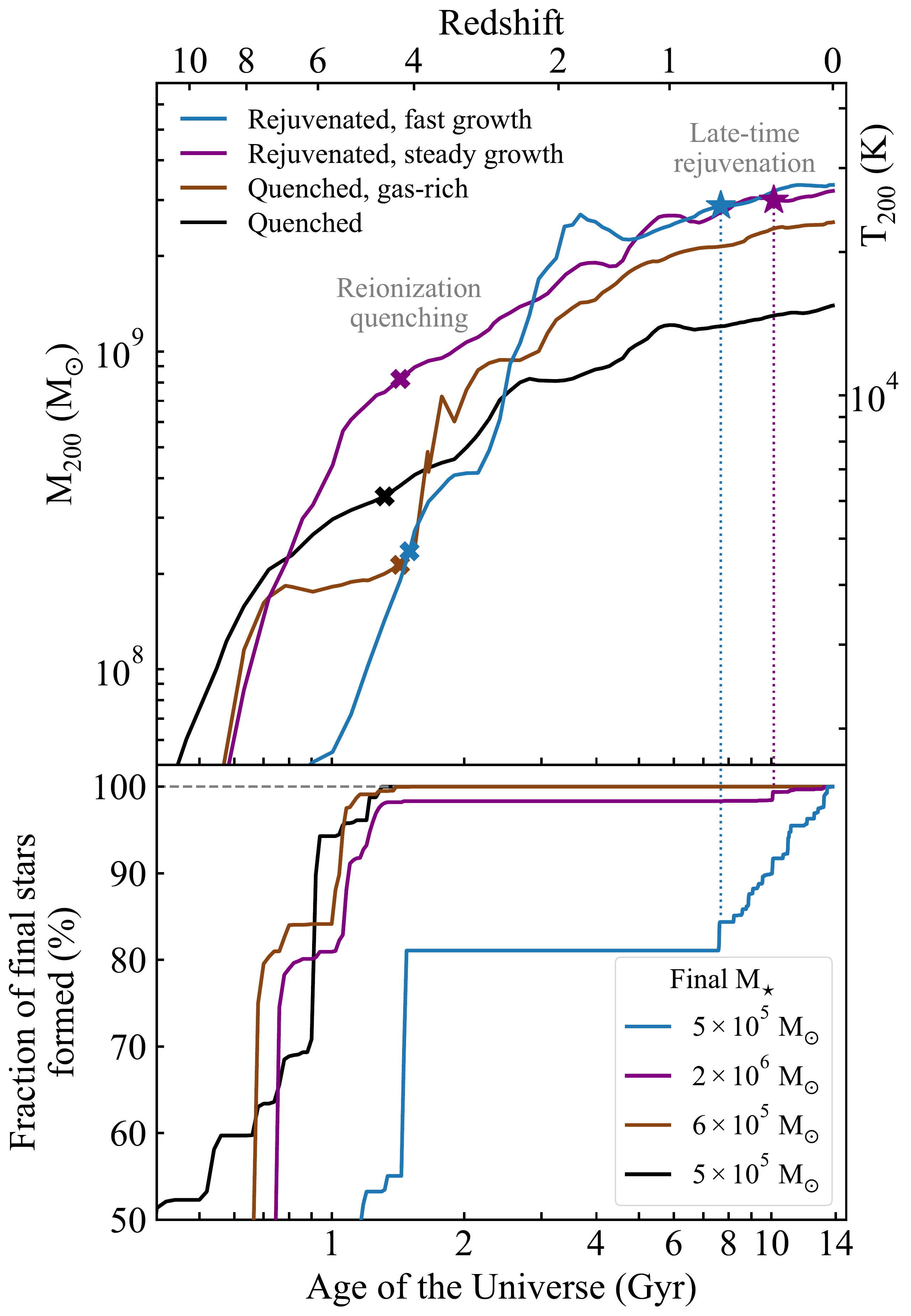}

    \caption{Growth in dynamical mass (top panel) for all galaxies in the simulated suite and their cumulative, archaeological star formation histories as would be observed in the Local Group (bottom panel). All galaxies show a halt in internal star formation at $z=4$, caused by the suppression of gas inflows following cosmic reionization at $z=6$. This halt is permanent for the two lower mass dwarfs (brown and black), which exhibit an uniformly old stellar population. In contrast, higher mass galaxies (blue and purple) are able to reignite star formation after $z=1$. Despite their virial temperatures (right axis) being close to the temperature of the IGM ($\simeq \xScientific{2}{4} \,$K), they are able to accrete gas at late times (Figure~\ref{fig:gasmass}), reform an interstellar medium, and sustain the formation of new stars until $z=0$. 
    }
    \label{fig:massgrowth}
\end{figure}

We select two high-mass haloes with contrasting histories, one dominated by major mergers at $z\approx3$ (blue in Figure~\ref{fig:massgrowth}) compared to a second with a steady mass growth (purple). We further select a low-mass halo (black\footnote{This galaxy was first presented in \citet{Rey2019b}, labelled as `Earlier'.}) with a high dynamical mass at the time of reionization but little mass growth thereafter. We complete our selection with an intermediate object (brown) in both history and final mass. 

In addition to these four reference initial conditions, we will create `genetically modified' variants for two galaxies in Section~\ref{sec:results:gms}. The genetic modification approach (\citealt{Roth2016, Rey2018}) generates a new initial condition for a galaxy, closely related to its reference, but introducing a targeted change to modify a specified aspect of the galaxy's mass accretion history. For example, in Section~\ref{sec:results:gms}, we will create modifications to increase the final halo mass of a galaxy (see also \citealt{Roth2016}) or modify its merger history at fixed halo mass (see also \citealt{Pontzen2017, Rey2019a, Rey2019b}). All untargeted features are maximally conserved between the new, modified initial condition and its original, ensuring, for example, that the large-scale structure around the galaxy remains the same (e.g. \citealt{Rey2019a}, fig. 1). These features allow us to construct controlled experiments, isolating the role of cosmological histories in shaping a galaxy's observables. We leave a detailed description of the modifications made in this work to Section~\ref{sec:results:gms}, as we wish to focus first on the four galaxies in our main suite.

\subsection{Galaxy formation physics}   \label{sec:setup:ramses}

We evolve all initial conditions to $z=0$ using cosmological zoomed galaxy formation simulations. We follow the evolution of dark matter, stars, and gas using the adaptative mesh refinement hydrodynamics code \textsc{ramses} (\citealt{Teyssier2002}). The dynamics of collisionless particles (dark matter and stars) are computed using a multiscale particle-mesh solver, estimating densities through a cloud-in-cell approximation (\citealt{Guillet2011}). Fluid dynamics are computed using an HLLC Riemann solver (\citealt{Toro1994}) and the fluid equations are closed by assuming an ideal gas equation of state with adiabatic index $\gamma = 5/3$. Our refinement strategy (see \citealt{Agertz2020}) allows us to reach a spatial resolution of $3 \, \pc$ throughout the galaxy's interstellar medium. We complement hydrodynamics with an extensive galaxy formation model described in detail by \citet{Agertz2020} as `Fiducial'. We briefly review the most important ingredients for this work and refer to \citet{Agertz2020} and Section~\ref{sec:results:physics} for a discussion of the sensitivity of our results to galaxy formation physics.

We track the cooling of a primordial plasma using hydrogen and helium equilibrium thermochemistry (last described in \citealt{Courty2004}, see also \citealt{Rosdahl2013}) accounting for photoionization, collisional ionization and excitation, recombination, bremsstrahlung, Compton cooling and heating, and dielectronic recombination. The contribution from metal lines to the total cooling is extracted from tabulated models generated with \textsc{cloudy} (\citealt{Ferland1998}). We model heating from reionization through a spatially uniform, time-dependent UV background as implemented in the public \textsc{ramses} version. This implementation is based on an updated version of \citet{Haardt1996}, complemented by a high-redshift cut-off sharply dropping the photoionization and heating rates to avoid overheating the IGM at early times ($z>10$; \citealt{Onorbe2017}). Using this fiducial model, the IGM reaches a temperature greater than $10^4\,$K at $z \simeq 6$ as expected for a late-reionizing region (e.g. \citealt{Keating2020}), consistent with our selection of isolated dwarf galaxies embedded in a cosmic void (Section~\ref{sec:setup:ics}). We will explore the sensitivity of our results to the timing of reionization in Section~\ref{sec:results:physics}.

Once the gas is allowed to cool, it collapses to reach densities high enough to self-shield against surrounding radiation. We include an on-the-fly self-shielding prescription (\citealt{Aubert2010, Rosdahl2012}), which exponentially damps the heating and ionization rates above a critical density threshold. The value of the threshold is fixed throughout this work to $n_{\text{H, crit}} = 0.01 \, \cmcube$ and was calibrated to reproduce neutral fractions obtained from radiative transfer calculations (e.g. \citealt{Pontzen2008, FaucherGiguere2010, Rosdahl2012}; see Appendix~\ref{app:shielding} for more details on the implementation).

Since gas cooling is increasingly efficient in denser regions, it proceeds as a run-away process and should lead to the formation of stars within a cooling time. Our simulations model star formation with a recipe following a Schmidt law:

\begin{equation}
  \label{eq:schmidt}
  \dot{\rho}_{*}=\epsilonff \frac{\rho_{g}}{\tff}\ \text{for gas cells with} \,\,\rho_{g}>\rho_{\star} \ \text{and} \ T_g < T_{\star} \, ,
\end{equation}
where $\dot{\rho}_{*}$ is the instantaneous star formation rate in a gas cell, $\epsilonff$ is the star formation efficiency per free-fall time, $\rho_{g}$ and $T_g$ are the gas cell density and temperature, $\tff=\sqrt{3\pi/32G\rho}$ is the local free-fall time and $\rho_\star$ and $T_{\star}$ are imposed thresholds that gas must satisfy to qualify for star formation. For every gas cell with density higher than $\rho_\star = 300 \, \mppercmcube$ and temperatures lower than $T_{\star} = 100$ K, we sample Equation~\eqref{eq:schmidt} stochastically through a Poisson process, ensuring that the mean number of stellar particles formed is proportional to $\dot{\rho}_{*}$ (\citealt{Rasera2006}). Our stellar particles have initial masses of $300 \ \Msol$ to ensure a complete sampling of the initial mass function (IMF; \citealt{Kroupa2001}).

A key aspect of our simulations is the modelling of feedback from massive stars, accounting for SNe II and Ia explosions and stellar winds from massive and asymptotic giant branch (AGB) stars. Our model tracks the stellar evolutionary time-scales of stars with different masses within a stellar particle (\citealt{Agertz2011}), ensuring that each feedback process is injected on its physically motivated time-scale. The budget of energy, momentum, mass, and metallicity returned to the interstellar medium by each mechanism is stated in \citet{Agertz2013}. 

We model SN explosions as discrete events, by computing at each simulation time-step the number of stars exiting the main sequence to turn into SNe II and Ia (equations 6 and 13 in \citealt{Agertz2013}). This IMF-averaged number is then randomly sampled through a Poisson process to obtain a discrete number of explosions (\citealt{Agertz2020}). Our resolution greatly reduces uncertainties in modelling the energy injection of SNe to the surrounding medium, being sufficient to capture the cooling radius of most individual SNe (\citealt{Kim2015, Martizzi2015}). This allows us to inject a thermal energy of $10^{51} \, \text{erg}$ for each resolved explosion and self-consistently follow the build-up of momentum by solving the hydrodynamics equations. We switch to momentum injection only when the cooling radius of SNe is unresolved. We also include the loss of mass due to AGB stars, continuously releasing the IMF-averaged mass-loss (eq. 17 in \citealt{Agertz2013}) over the lifetime of a stellar population.

Despite the accurate modelling of SN feedback in our fiducial model, uncertainties on the star formation histories and stellar masses of simulated ultrafaints remain. Additional feedback channels can weaken or strengthen the coupling of SN explosions to the interstellar medium of dwarf galaxies, modulating its efficiency (e.g. \citealt{Smith2019, Agertz2020} -- see also \citealt{Munshi2019} for an exploration of alternative star formation prescriptions). In Section~\ref{sec:results:physics}, we will discuss the sensitivity of our results to such residual uncertainties, introducing re-simulations of our base suite with alternative galaxy formation models.

We compute the build-up of mass and merger trees of galaxies using the \textsc{pynbody} (\citealt{Pontzen2013}) and \textsc{tangos} (\citealt{Pontzen2018}) libraries, matching haloes between simulation snapshots based on their dark matter particle IDs. Cooling times quoted throughout the paper are obtained by evaluating the internal cooling function of \textsc{ramses} as a function of gas density, temperature, and metallicity. \MR{We interpolate a single stellar population model (\citealt{Girardi2010}) over a grid of ages and metallicities to obtain the luminosities of individual stellar particles.} We then sum them to obtain total absolute $V$-band magnitudes, and pick a random line of sight to derive the projected half-light radii quoted in Table~\ref{table:runs}.

\section{Results} \label{sec:results}

We now describe the evolution in gas content and star formation across our suite of dwarf galaxies. We focus first on the four galaxies in our main suite, showing that the two most massive objects are able to reignite star formation at late times (Sections~\ref{sec:results:sf} and~\ref{sec:results:cooling}). We then use genetic modification to show that this late reignition is controlled by the interplay between residual stellar feedback from SNe Ia and AGB stars (Section~\ref{sec:results:residualFB}) with a dwarf's specific mass accretion history (Section~\ref{sec:results:gms}). We finally explore the sensitivity of our results to residual uncertainties in our galaxy formation model (Section~\ref{sec:results:physics}).

\subsection{Late-time star formation in low-mass dwarfs} \label{sec:results:sf}

Figure~\ref{fig:massgrowth} shows the growth in dynamical mass for each of the four galaxies in the suite (top panel) and their cumulative, archaeological star formation history (bottom panel) as would be derived from observing a colour--magnitude diagram of the galaxy in the Local Group today (e.g. \citealt{Weisz2012, Brown2014}). We select all stars in the galaxy at $z=0$ and compute \MR{their fractional, cumulative star formation histories}. The final stellar masses of each galaxy are quoted in the legend of the bottom panel. 

All four galaxies exhibit a plateau in star formation after $z=4$, indicative of a halt in internal star formation. We mark by a cross the \MR{time at which the last old stellar population is formed across all progenitors of the final galaxy. Despite} having an order-of-magnitude difference in dynamical mass at $z = 4$, all galaxies quench at nearly identical times. This global suppression of star formation is a signature of cosmic reionization, heating the IGM and preventing inflows of fresh gas onto these small galaxies. Reionization is completed by $z\simeq6$, but high-density, self-shielded gas within the halo allows star formation to continue (e.g. \citealt{Susa2004, Onorbe2015}). The suppression of inflows prevents this gas reservoir from being replenished, leading to the quenching of \textit{in situ} star formation $500$ Myr later ($z\simeq4$; e.g. \citealt{Wheeler2019, Agertz2020}). This delay is consistent with the observed ages of stars in quenched ultrafaint dwarf galaxies (e.g. \citealt{Okamoto2012, Brown2014, Weisz2014}).

By $z = 4$, the `Quenched' and `Quenched, gas-rich' galaxies (respectively, black and brown) have formed 100 \MR{per cent} of their stars, exhibiting an uniformly old stellar population today. In contrast, the `Rejuvenated, fast growth' and `Rejuvenated, steady growth' galaxies (blue and purple) experience late-time star formation, forming, respectively, 19 and 2 \MR{per cent} of their stars after $z=1$. Following their reignition (marked by a star in the top panel), both galaxies form new stellar populations at an average star formation rate of $\xScientific{1.5}{-5} \, \Msolyr$ and $\xScientific{0.9}{-5} \, \Msolyr$, respectively. Our results demonstrate that low-mass dwarf galaxies can reignite star formation after quenching by cosmic reionization, and are then able to \textit{sustain} extremely low star formation rates ($\sim 10^{-5} \, \Msolyr$) over several billions of years. 

The differing behaviours of star formation histories after cosmic reionization can be explained at first order through the interaction between the growth history of each galaxy and a `threshold' dynamical mass that has to be reached to accrete gas from the IGM after reionization (e.g. \citealt{Efstathiou1992, Gnedin2000, Hoeft2006, Okamoto2008, Noh2014, BenitezLLambay2015, Fitts2017, Ledinauskas2018}). To visualize this interaction, we compute the time evolution of each galaxy's virial temperature (right axis in Figure~\ref{fig:massgrowth}) with
\begin{equation}
  \Tvir = \frac{\mu \, \protonmass \, G \, \Mvir}{2 \, k_B \, \rvir} \, ,
\end{equation}
where $\mu=0.59$ is the mean molecular weight of ionized primordial gas, $\protonmass$ is the proton mass, and $G$ and $k_B$ are the gravitational and Boltzmann constants, respectively. If $\Tvir \ll \Tigm$, the potential well of the galaxy is too shallow to overcome thermal pressure and gas accretion is prevented. 

Reionization heats the IGM to $\Tigm \simeq 2\times10^4 \,$K, i.e. above the corresponding virial temperature of all galaxies at $z = 4$, thus consistent with the quenching of internal star formation. At later times, three galaxies are able to grow sufficiently in dynamical mass to fulfill $\Tvir \gtrsim \Tigm$. However, despite all three galaxies having sufficiently deep potential wells, only two are forming stars at late times, showing that the the sole interplay between mass growth history and IGM temperature does not provide a complete picture of the ability of these galaxies to reignite their star formation. 

In particular, the narrow spread of our galaxies in virial temperature makes our argument particularly sensitive to the precise value of $\Tigm$, when, in fact, the thermal state of IGM gas is both density and time dependent (see \citealt{McQuinn2016_review} for a review on the thermal evolution of the IGM). For example, the reionization of helium by the population of quasars at $z \approx 3$ further heats the IGM, although the overall temperature increase remains uncertain (e.g. \citealt{McQuinn2009, Compostella2013}). Even when incorporating such density and time dependences, arguments based on the virial temperature still neglect the gas physics in the interior of haloes, and the importance of astrophysical effects such as feedback. We will see in the next sections that these processes are key to determine the ability for a dwarf to reignite star formation -- we focus on them first and defer a discussion of the importance of IGM physics to Section~\ref{sec:results:physics}.

We finally note that the star formation rates and histories of the two higher mass galaxies provide a good match to those of low-mass dwarfs Leo~T (\citealt{Clementini2012, Weisz2012}) and Leo~P (\citealt{McQuinn2015LeoP}). We leave a more detailed observational comparison of the simulated and observed \hi \ kinematics and abundance patterns to a follow-up work, and focus next on pinpointing the physical processes driving the reignition of star formation and their interaction with the mass growth history of dwarf galaxies.

\subsection{The reignition of star formation} \label{sec:results:cooling}

\begin{figure}
  \centering
    \includegraphics[width=\columnwidth]{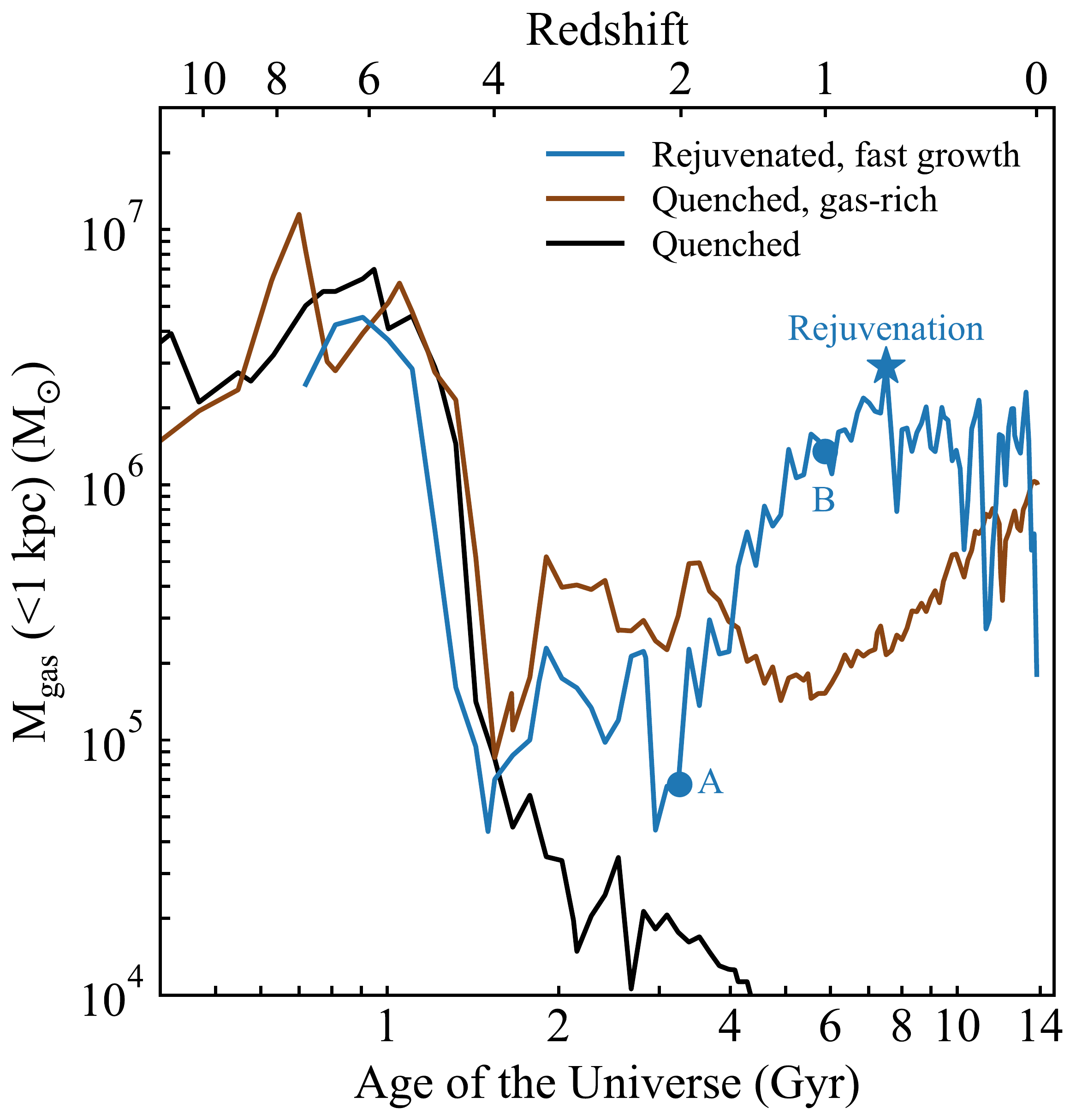}

    \caption{Evolution of the central gas mass for three low-mass dwarfs with increasing final halo mass. Reionization suppresses the gas content of all galaxies by an order of magnitude between $z = 7$ and $4$, but the two more massive galaxies (blue and brown) are able to retain gas in their centre. The most massive example (blue) then sees its central gas content steadily increase from $z\approx2$, ultimately leading to the rejuvenation of star formation. We expand the gas phase diagrams at the beginning (A) and end (B) of the transition (Figure~\ref{fig:gasphase}) to show that accretion is driven by a slow, nearly isothermal collapse of gas. The intermediate object (brown) also presents an increase in central gas mass, although delayed in time. We will show that this system is close to, but has not yet completed the reignition of star formation (Figure~\ref{fig:agbs}).
    }
    \label{fig:gasmass}
\end{figure}

Figure~\ref{fig:gasmass} shows the evolution of the central gas mass for three dwarf galaxies with increasing final dynamical mass. We only present a single rejuvenating galaxy (blue) for visual clarity as the second (`Rejuvenated, steady growth') exhibits a similar evolution in gas mass and gas phase. We compute the central gas mass by constructing the spherically averaged, enclosed gas mass for each galaxy at $1 \, \kpc$. This radius is chosen to enclose the central region around the galaxy, well within the virial radius ($\sim 10 \, \kpc$), while enclosing several stellar half-light radii ($\sim 0.1 \, \kpc$) at all times. We verified that the trends observed in Figure~\ref{fig:gasmass} are unchanged when computing the gas mass at $0.1$, $0.5$, and $2.0\, \kpc$ \MR{in physical and comoving units}.

We first observe in Figure~\ref{fig:gasmass} the strong impact of reionization between $z = 7$ and $4$, cutting off gas inflows and leading to a decrease of central gas content in all galaxies by an order of magnitude. Gas ejected by the last SN outflows cannot be replenished from the recently heated IGM, leading to this strong reduction in gas content. After $z = 4$, however, evolutions start diverging. The more massive galaxies (brown and blue) are able to retain significantly more gas in their centre compared to the lowest mass galaxy (black). 

Furthermore, starting from $z \approx 2$, the most massive galaxy (blue) sees its central gas content steadily increase, signaling ongoing accretion driving gas towards the galaxy's centre. Steady accretion of gas refuels its interstellar medium, eventually leading to sufficiently high densities to ignite a burst of star formation. Following this initial event, the central gas content remains roughly constant on $\Gyr$-time-scales -- visual inspection shows that peaks in gas mass coincide with star formation bursts, indicative of a regulating cycle between gas accretion and feedback. 

We also observe in Figure~\ref{fig:gasmass} that the intermediate-mass galaxy (brown) undergoes a similar increase in central gas content to its higher mass counterparts, although delayed in time. By $z=0$, this galaxy has not reached as high gas masses as the two more massive counterparts, consistent with its absence of late-time star formation (Figure~\ref{fig:massgrowth}). Because of its similar evolution but lower dynamical mass, we hypothesize that this system is undergoing gas accretion, but has not yet completed the rejuvenation of star formation by $z=0$. We will show that the reignition of star formation can indeed be forced in this system by either decreasing gas heating from feedback in its centre (Section~\ref{sec:results:residualFB}) or increasing its dynamical mass (Section~\ref{sec:results:gms}) using a genetic modification.

\begin{figure}
  \centering
    \includegraphics[width=\columnwidth]{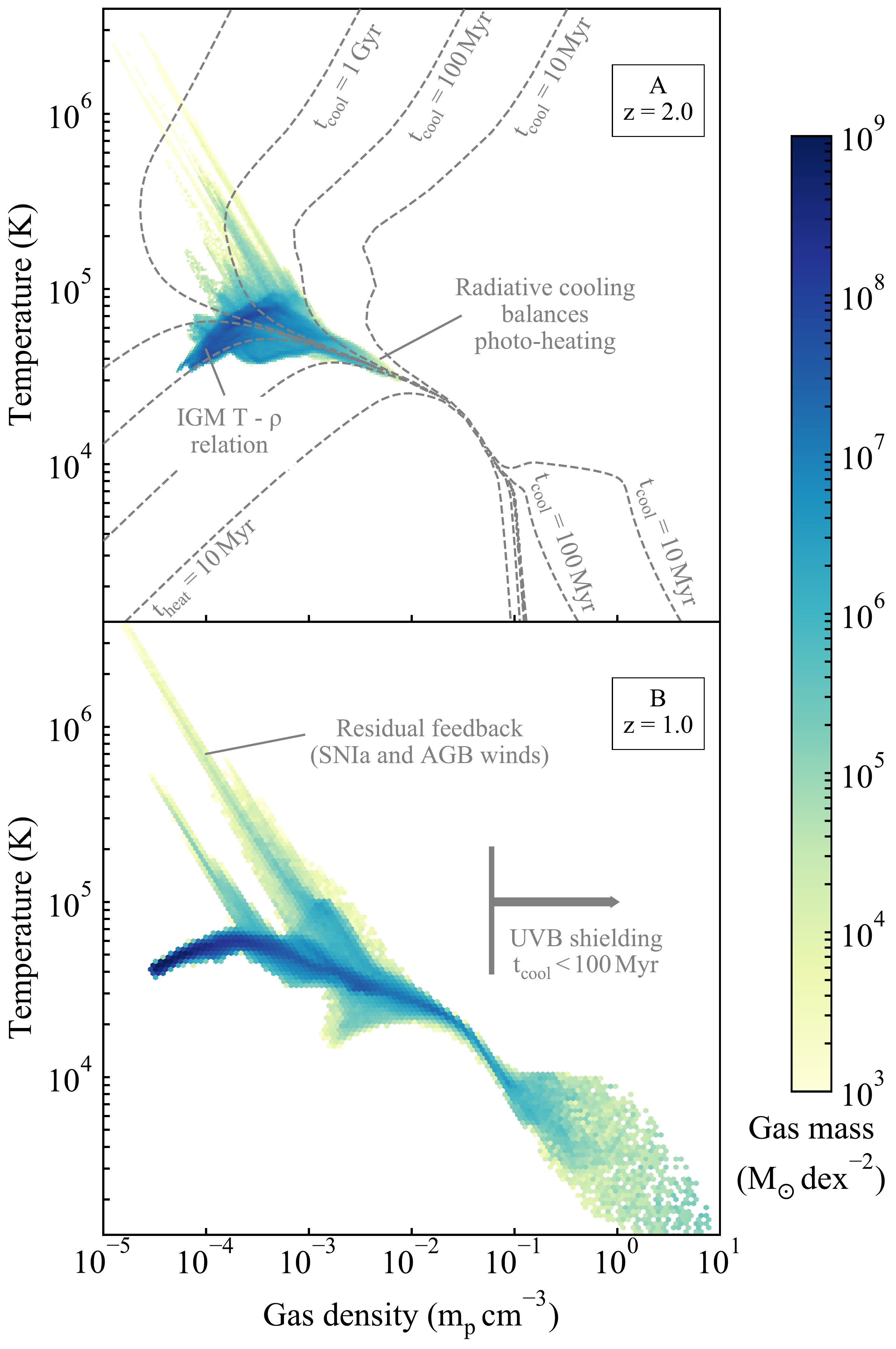}

    \caption{Distribution of gas temperatures and densities during the accretion preceding rejuvenation. At the beginning of the transition (top panel), gas is distributed along the temperature--density relation of the IGM at low densities, and the equilibrium between radiative cooling and photoheating at higher densities. Gas compresses along this equilibrium, reaching sufficient densities to self-shield against the surrounding UV background towards the end of the transition (bottom panel). Once shielded, gas has short cooling time ($\lesssim 100 \, \Myr$) and should rapidly lose thermal support to fuel star formation.  Discrete plumes of high-temperature gas are however visible in both panels, signaling active residual feedback from SNe Ia and AGB stars. These heating sources are sufficient to balance gas cooling, delaying the onset of star formation by several billion years (Figure~\ref{fig:agbs}).}
    \label{fig:gasphase}
\end{figure}

To gain insights into the physical mechanisms driving the increase in gas content, we select two snapshots at the beginning and end of the transition (A and B in Figure~\ref{fig:gasmass}, respectively) and extract the corresponding gas-phase diagrams (Figure~\ref{fig:gasphase}). We select all gas within the virial radius and compute two-dimensional logarithmic bins in gas temperature and density. The colour of each bin shows the sum of cell gas masses with this temperature and density per bin area. We further show contours of constant cooling and heating times (grey dashed), to gain intuition into the evolutionary time-scales of gas.

At the start of gas accretion ($z=2$, top panel), all gas is hotter than $10^4\,$K, and can be split into two regime. At low densities ($\rho \lesssim \xScientific{3}{-4} \, \mppercmcube$), gas in the first regime follows the relationship between temperature and density induced by out-of-equilibrium photoheating in the IGM ($T \propto \rho^{\gamma-1}$ with $\gamma\sim1.55$; \citealt{Theuns1998, Gnedin2000, McQuinn2016_TIGM}). Some scatter around the mean relation is also observed, as expected if gas undergoes shock and dynamical heating during non-linear collapse. This feature consists of gas being heated by the UV background, on time-scales greater than a billion year (bottom left-hand contours), thus unlikely to condense and reach the halo centre. A second regime at higher densities ($\rho \gtrsim \xScientific{3}{-4} \, \mppercmcube$) is also apparent. Once sufficient densities are reached to radiate energy efficiently, the gas follows a new equilibrium regime set by the balance between radiative cooling and photoheating. As can be seen from the convergence of heating and cooling times contours, the equilibrium temperature is nearly isothermal with increasing densities (\citealt{Theuns1998, Noh2014}). If able to condense, gas following this equilibrium would produce a nearly isothermal gas density profile, leading to large gas overdensities in the centre (\citealt{Ricotti2009}).

When reaching the end of the transition ($z=1$, bottom panel), most of the halo gas is distributed along the equilibrium between photoheating and radiative cooling and has evolved to higher densities. We also observe a `knee' in the equilibrium temperature around gas densities $\gtrsim 10^{-2} \, \mppercmcube$, also visible in the top-panel contours. The transition is a consequence of self-shielding (Section~\ref{sec:setup:ramses}), which reduces the impact of photoheating above this density threshold, allowing gas to cool below $10^4\,$K.

We therefore find that gas accretion and the rejuvenation of star formation at late times occurs through a nearly-isothermal condensation extending over several billion years. Gas condenses along the equilibrium between radiative cooling and photoheating, until sufficient densities are reached to self-shield from the UV background and fuel star formation. We verified that the phase evolution observed in Figure~\ref{fig:gasphase} is generic across all galaxies exhibiting gas accretion. However, this physical picture leaves several questions which we address now.

\textit{What drives gas towards higher densities?} In Figure~\ref{fig:gasphase}, gas along the equilibrium temperature is in thermal balance, neither losing or gaining thermal support, and needs to increase in density before star formation can occur. External factors (mergers or environmental interactions) can compress a galaxy's interstellar medium, but (i) we verified that our dwarfs have no visible encounters with surrounding gas structures (e.g. \citealt{Wright2019}) and (ii) both galaxies with contrasting merger histories (rapid major mergers and steady growth) exhibit a reignition of star formation (Figure~\ref{fig:massgrowth}). Furthermore, Figure~\ref{fig:gasmass} shows that 5 billion years elapse between the beginning of gas accretion and the reignition of star formation, which would be inconsistent with a rapid, external triggering event. We argue that this condensation of gas is rather linked to the slow build-up of dynamical mass at late times, which we explore further in Section~\ref{sec:results:gms}.

\textit{Why is the transition between self-shielding and star formation delayed?} As shown by the contours in Figure~\ref{fig:gasphase}, self-shielded gas should cool rapidly, in at most $100 \, \Myr$. High-density gas at $z=1$ (bottom panel in Figure~\ref{fig:gasphase}) should therefore collapse, driving a cooling flow that should rapidly lead to the reignition of star formation. This is in stark contrast with the observed $1.5 \, \Gyr$-delay between $z=1$ and the reignition of star formation (Figure~\ref{fig:gasmass}). 

However, the quoted cooling times assume that the UV background is the only heating source in our galaxies, when, in fact, residual feedback from SNe Ia explosions and AGB winds is active long after the quenching of internal star formation by reionization. Both panels of Figure~\ref{fig:gasphase} exhibit discrete plumes of high-temperature gas, which we tracked to individual bubbles of hot gas expanding in the galaxy's interstellar medium. The observed scatter in gas temperature at high, self-shielded gas densities (bottom panel of Figure~\ref{fig:gasphase}) is further evidence of active feedback from old stellar populations and we focus next on determining their role in regulating the balance between cooling and heating.   

\subsection{Residual feedback delays the reignition of star formation} \label{sec:results:residualFB}

We now demonstrate that the combined activity of SNe Ia and winds from AGB stars inject sufficient energy to delay the reignition of star formation by several billion years. To achieve this goal, we focus on the intermediate-mass galaxy, as we postulated in Section~\ref{sec:results:cooling} which it is close to rejuvenating star formation.

\begin{figure}
  \centering
    \includegraphics[width=\columnwidth]{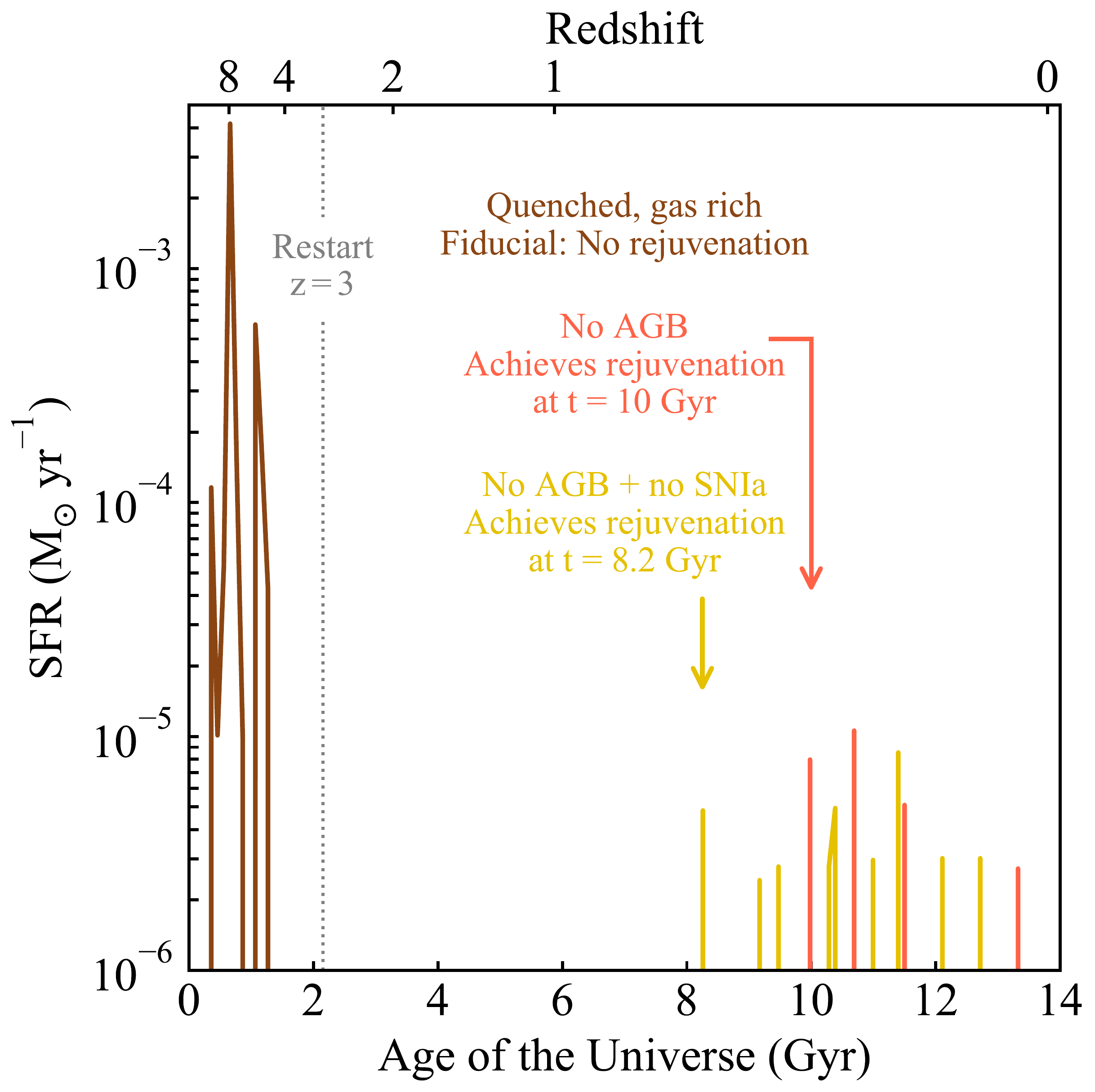}

    \caption{The sensitivity of star formation histories to residual feedback from old stellar populations. Figure~\ref{fig:gasmass} shows that the intermediate-mass galaxy (brown) is undergoing gas accretion and is likely close to rejuvenate. Restarting the run at $z=3$ and removing the contribution of AGB stars (red) to the feedback budget is able to force rejuvenation in this system. Further removing the contribution of SNe Ia (yellow) quickens the reignition, demonstrating that these two residual feedback sources are delaying the onset of star formation by several billion years.}
    \label{fig:agbs}
\end{figure}

We re-start from our fiducial run at $z=3$, removing any injection of mass, momentum, and energy first from stellar winds alone, and then from both winds and SNe Ia explosions. We caution that such experiments are by construction unphysical, but they are useful in isolating the importance of residual feedback from old stellar populations. We show the resulting star formation histories in Figure~\ref{fig:agbs}. As previously described in Section~\ref{sec:results}, the fiducial galaxy (brown) shows no star formation activity past $z=4$. Removing the contribution of stellar winds to the feedback budget (red) allows this system to rejuvenate star formation when the Universe is $\approx 10 \, \Gyr$ old. Further removing the contribution of SNe Ia (yellow) brings the rejuvenation of star formation even earlier, at $8.2 \, \Gyr$, allowing the galaxy to build 2 \MR{per cent} of its final \MR{stars} at late times at an average star formation rate of $\xScientific{2.2}{-6} \, \Msolyr$. Comparing these galaxies, we conclude that the injection of energy from the combination of SNe Ia and stellar winds is sufficient to balance the cooling of gas in such low-mass systems, delaying a potential reignition of star formation by a significant fraction of the Hubble time (here as much as $6 \, \Gyr$).

The precise value of this delay will depend on both the efficiency of residual feedback and the dynamical mass of a given dwarf. With our currently implemented model, the strongest contributor to the delay are AGB stars. However, our model approximates the injection of stellar winds in the interstellar medium through a continuous, IMF-averaged--release of mass over the lifetime of a stellar population. New insights could be gained by, for example, discretely sampling AGB stars as is currently done for SNe or refining the momentum and energy injection into the interstellar medium from the unresolved AGB bow shock (see the discussion in, e.g. \citealt{Conroy2015}). A systematic study, which we leave for future work, should also explore uncertainties in the rate of SNe Ia, and its time evolution through the delay--time distribution (e.g. \citealt{Mannucci2006, Maoz2012}). 

This study is sufficient to establish that, in small dwarf galaxies, low levels of feedback from old stellar populations can be sufficient to balance the cooling of gas for a significant fraction of the Hubble time. In addition to revisiting implementations of SNe Ia and winds, future studies aiming for a detailed understanding of residual feedback in low-mass dwarfs will have to understand its coupling with each dwarf's mass accretion history, which is the focus of the next section.

\subsection{The coupling between rejuvenation and the mass build-up of dwarfs} \label{sec:results:gms}

In Section~\ref{sec:results:cooling}, we argued that the slow condensation of gas is linked to the late build-up of a dwarf's dynamical mass. Such condensation can then be balanced by residual feedback from old stellar populations (Section~\ref{sec:results:residualFB}). In this section, we probe this interplay between mass growth and feedback using our ability to create alternative mass accretion histories for our dwarf galaxies (see also \citealt{Rey2019b}). 

\begin{figure*}
  \centering
    \includegraphics[width=\textwidth]{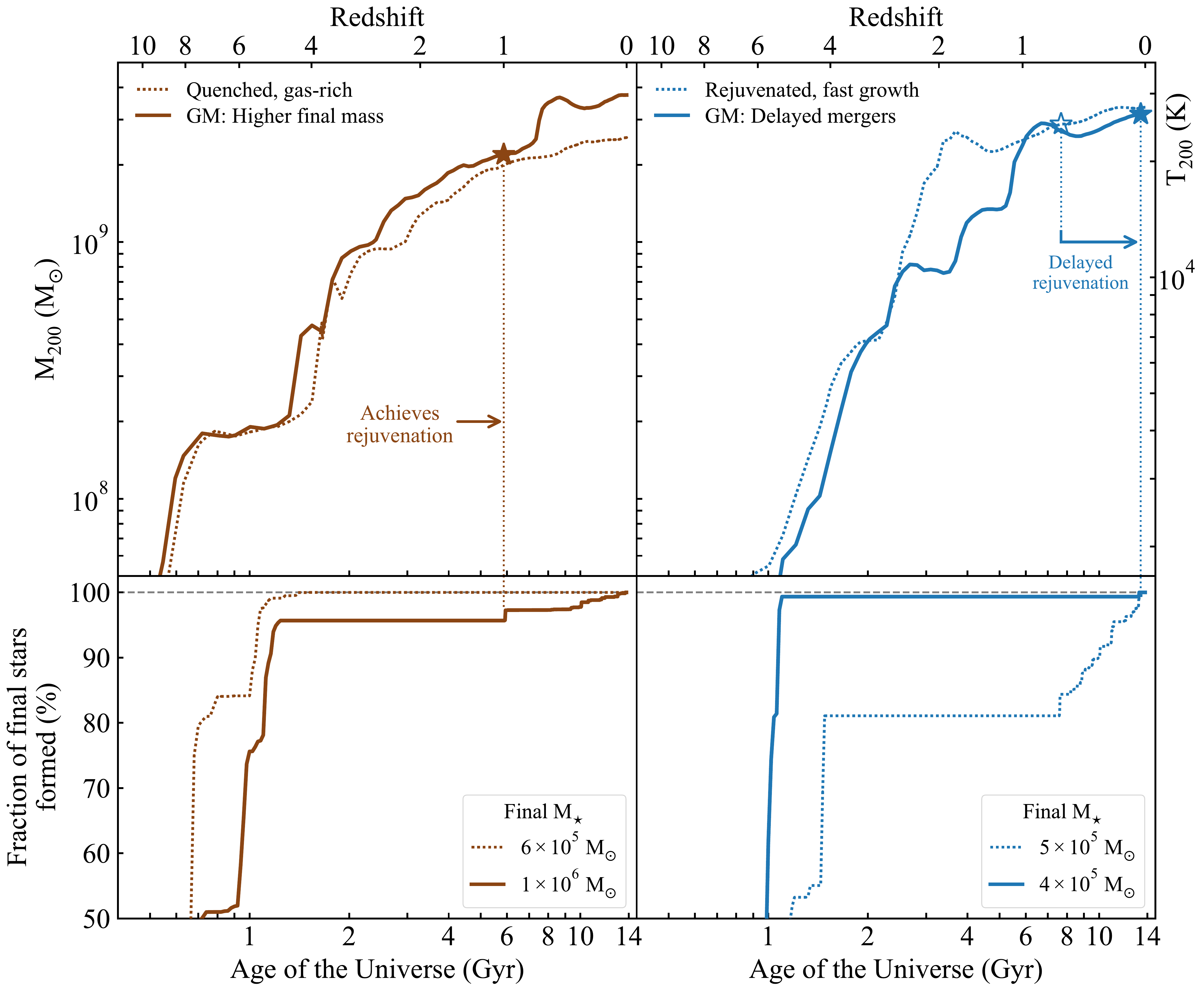}

    \caption{Impact of modifying the mass build-up of dwarf galaxies (top panel) on their star formation (bottom panel). Using the genetic modification approach, we modestly increase the late-time dynamical mass of our intermediate object (brown, left-hand panel). It forces the reignition of star formation at $z \approx 1$, when the reference galaxy (dotted) remains quiescent up to $z=0$. We also delay the delivery of mass in a rejuvenating galaxy (blue, right-hand panel), which delays the reignition of star formation, nearly preventing it altogether compared to the reference case (dotted). The extended time-scale required to reignite star formation therefore strongly interacts with the growth history of dwarf galaxies, and will lead to a diversity in their stellar ages and gas content.
    }
    \label{fig:gms}
\end{figure*}

\subsubsection{Forcing rejuvenation by increasing halo mass}
We start by constructing a genetically modified initial condition for the intermediate-mass galaxy (`Quenched, gas-rich'). We have shown in Section~\ref{sec:results:residualFB} that rejuvenation in this system can be forced by reducing internal heating from residual feedback. We wish to test whether the same fate can be forced by increasing its dynamical mass and so construct a modified initial condition designed to increase its halo mass at late times. The final halo mass can be controlled by altering the mean density across the Lagrangian region (\citealt{Roth2016}) -- we therefore track back to the initial conditions all particles within the galaxy at $z=0$ and increase the mean density within this region by 20 \MR{per cent}. 

Figure~\ref{fig:gms} (left-hand panel) shows the reference and modified mass growth histories (dotted and solid brown respectively). The modified mass growth is consistently higher at late times compared to the reference, reaching $\xMsol{3.7}{9}$ at $z=0$ compared to $\xMsol{2.5}{9}$. This increase in final dynamical mass has two effects on the central galaxy: (i) an increase in the final stellar mass from $\xScientific{6}{5}$ to $\xMsol{1}{6}$ and (ii) a reignition of star formation at $z=0.96$, with 5 \MR{per cent} of the \MR{final stars} formed at late times, at an average star formation rate of $\xScientific{9.6}{-6} \, \Msolyr$. The increase in final stellar mass is primarily driven by the accretion of old stars previously formed in nearby ultrafaints, rather than to the production of new \textit{in situ} stars. In particular, a late merger at $z=0.82$, visible as a sudden increase in dynamical mass (top panel), provides an additional $\xMsol{5}{5}$ of stellar mass to the final, modified galaxy.

With this experiment, we show that a modest increase in dynamical mass is capable of forcing rejuvenation in a previously quiescent system. We also showed in Section~\ref{sec:results:residualFB} that reignition could be forced in this system by reducing the energy input from residual feedback. Combining these two results, we conclude that the barrier to gas accretion and star formation opposed by residual feedback can be overcome if the galaxy undergoes more active mass build-up at late times.

\subsubsection{Delaying rejuvenation by delaying mergers}
We now construct a genetically modified initial condition starting from the `Rejuvenated, fast growth' galaxy to see whether its rejuvenation can be delayed. The mass growth history of this object is dominated by a rapid series of major mergers around $z\sim3$. We wish to test whether delaying the build-up of mass due to these mergers would delay the reignition of star formation, at fixed halo mass today. Controlling multiple mergers is best achieved by modifying the local variance in the Lagrangian region (\citealt{Rey2018}). We follow the procedure described in detail by \citet{Rey2019a}, reducing the variance in the Lagrangian region by 20 \MR{per cent} on the scale matching the dynamical mass of the major progenitor at the time of the mergers (i.e. $\xMsol{1}{9}$). We also maintain the same mean density across the Lagrangian region to conserve the final halo mass (\citealt{Pontzen2017, Rey2019a}). 

Figure~\ref{fig:gms} (right-hand panel) shows the reference (dotted blue) and modified (solid) mass growth histories. Our modifications delay the mergers around $z=3$, as the reference galaxy grows from $\xScientific{4}{8}$ to $\xMsol{2}{9}$ in  $\approx 1.5 \, \Gyr$ from $z=3$ while the modified counterpart undergoes the same growth in $\approx 4 \, \Gyr$. The early growth and final halo mass from the reference galaxy are reproduced in the modified case, as expected from our modifications and the minimal nature of the changes operated by the genetic modification algorithm (\citealt{Rey2019a}). Delaying the mass build-up however has a radical impact on the cumulative star formation history (bottom panel). Only 0.7 \MR{per cent} of \MR{stars are} formed after $z=4$ compared to 19 \MR{per cent} in the unmodified history, and the reignition of star formation is delayed from $z=0.67$ to $0.04$ (arrow). The final stellar mass in the modified galaxy is slightly reduced, due to the shortfall of late-time stars. With this second experiment, we have shown that, at fixed halo mass today, galaxies building up dynamical mass more slowly will rejuvenate star formation later. 

Given the variety of possible histories allowed for a galaxy in a Lambda cold dark matter universe, our results therefore predict a wide diversity in stellar ages and gas content for low-mass dwarf galaxies. The interplay between each dwarf's mass growth and the combined feedback from the UV background and old stellar populations naturally predicts (i) field dwarfs galaxies that have grown sufficiently in mass to have ongoing star formation at low star formation rates, and (ii) field dwarfs `caught in transition', which have started to pile-up gas but have not built-up sufficient dynamical mass to reignite star formation. 

We note that observational evidence might already be present for individual examples belonging to each class of systems, i.e. gas-rich, star-forming dwarfs (e.g. Leo~T, Leo~P; \citealt{Irwin2007, McQuinn2015LeoP}), and gas-rich, low-mass dwarfs with uniformly old stellar populations (e.g. \citealt{Janesh2019}). A detailed prediction of the abundance of each class of systems requires a larger suite of objects, to quantify the sensitivity to final dynamical mass and growth history. This will be important to interpret the combination of next-generation \hi \ and imaging surveys which will directly probe the formation of low-mass dwarf galaxies in the field and is left for future work (Kim et al. in preparation; see also \citealt{Tollerud2018}).

\begin{figure}
  \centering
    \includegraphics[width=\columnwidth]{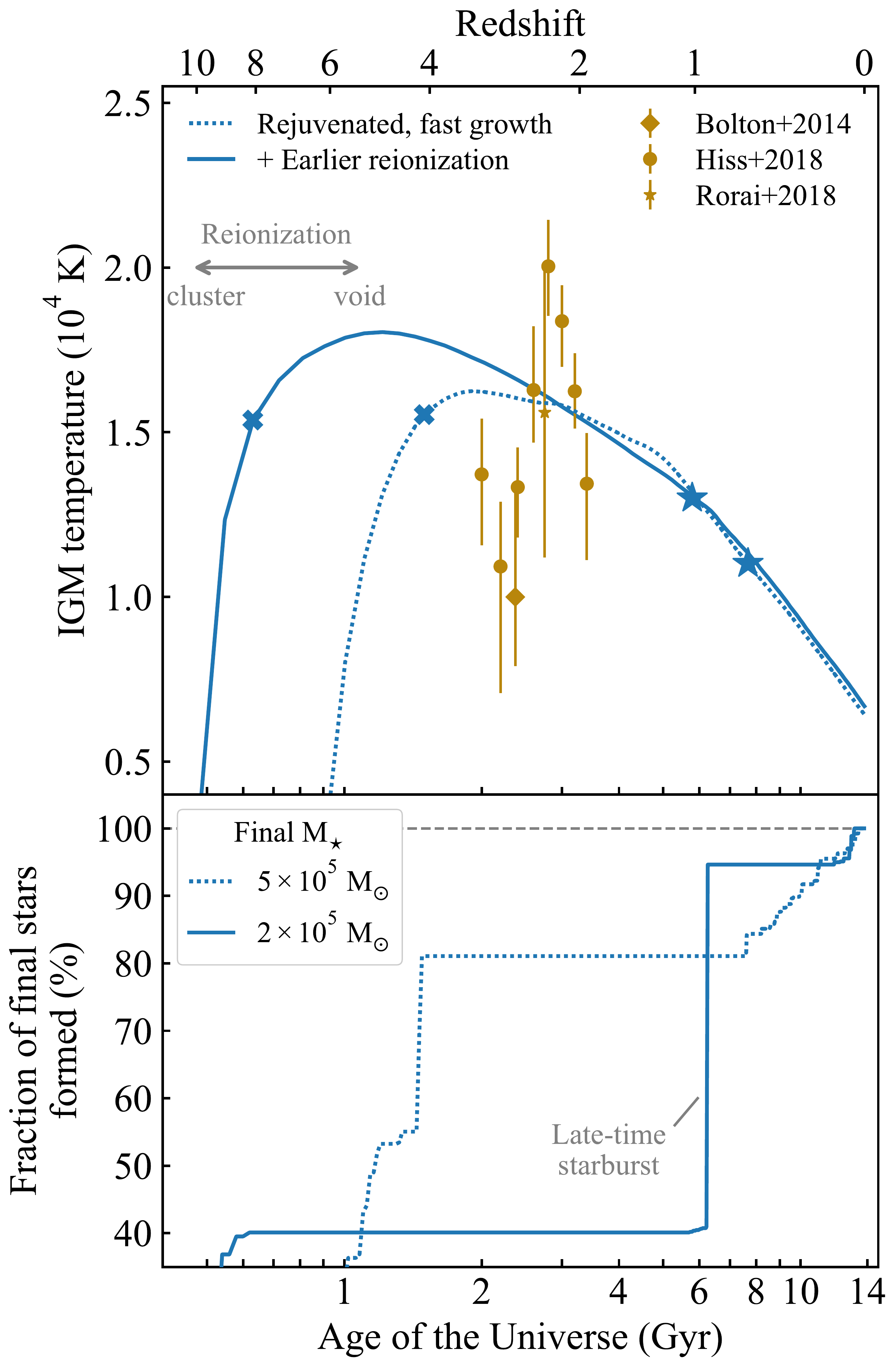}

    \caption{Impact of modifying the timing of hydrogen reionization on the star formation history of dwarf galaxies (bottom panel). With an earlier reionization (blue), the IGM is heated above $10^4\,$K at $z \approx 9$ (top panel), as would be expected in a denser environment than our fiducial cosmic void (dotted). Internal star formation is in turn quenched earlier at high redshift, but still reignites at late times to form a nearly identical amount of young stars than in the fiducial case. The timing of reionization is therefore imprinted in the final stellar mass (legend) and the old to young proportion of stars, but does not impact late-time star formation. \MR{Both thermal histories are consistent with observational constraints after reionization (gold).}}
    \label{fig:reionization}
\end{figure}

\subsection{Exploring residual modelling uncertainties} \label{sec:results:physics}

In the previous sections, we have established how the interplay between the growth of dynamical mass at late times and heating from both the UV background and old stellar populations controls the reignition of star formation. In this section, we test the sensitivity of this process to uncertainties in our numerical modelling.

\subsubsection{The timing of reionization} \label{sec:results:physics:reionization}
By heating the surrounding IGM, cosmic reionization is a key driver in quenching star formation at high redshift in our dwarf galaxies. Large uncertainties however remain in the average UV background and associated thermal history of the IGM at such times (e.g. \citealt{Onorbe2017}). Furthermore, reionization is, by nature, inhomogeneous -- overdense regions reionize their surroundings first, potentially as early as $z = 15$ (e.g. \citealt{Aubert2018}), while voids are reionized last, potentially as late as $z = 5.5$ (e.g. \citealt{Keating2020}). This, in turn, introduces an environmental modulation, varying the timing of reionization for a given dwarf galaxy (e.g. \citealt{Aubert2018, Ocvirk2018, Katz2019}). 

A self-consistent treatment of inhomogeneous reionization places conflicting demands on available resources, as it requires capturing both large, cosmological scales and the interstellar medium within small dwarfs (e.g. \citealt{Iliev2014, OShea2015, Pawlik2017, Finlator2018, Ocvirk2018, Rosdahl2018}). This study focusses on the latter, but we can nonetheless gain insights in the impact of spatial variation by varying the timing of reionization due to the uniform UV background. To this end, we re-evolve our prototypical rejuvenating galaxy (`Rejuvenated, fast growth') to $z=0$, removing the high-redshift cut-off in our fiducial UV background to force an earlier reionization. We visualize the induced change in thermal history by fitting the temperature--density relation of the IGM (following the procedure in \citealt{Onorbe2017}) and showing the evolution of the temperature at the mean cosmic baryon density in Figure~\ref{fig:reionization} (top panel). We compare these resulting thermal histories with observational determinations (\citealt{Bolton2014, Hiss2018, Rorai2018}).

Our reference model (dotted) ensures that the IGM is heated to temperatures $\geq 10^4 \,$K at $z \approx 6$, which corresponds to a late reionization. This is consistent with our selection of isolated dwarf galaxies, deep in a cosmic void. Removing the damping of photoionization and heating rates at high redshift (blue) causes reionization at $z \approx 9$, closer to what would be expected in an average density, non-void environment (\citealt{Aubert2018}). The two thermal histories match after reionization and are broadly consistent with the range allowed by observational data. 

As expected, earlier reionization leads to an earlier quenching of internal star formation, shifting from $z \approx 4$ to $\approx 7$ (crosses). Despite this earlier quenching, the galaxy undergoing earlier reionization still exhibits reignition and star formation at late times (bottom panel), forming $58$ \MR{per cent} of its stars after $z=1$. The average star formation rate at late times, $\xScientific{1.2}{-5} \, \Msolyr$, is nearly identical to the reference case, though a large fraction is produced in a starburst shortly after rejuvenation (labelled) and therefore might be subject to stochasticity (\citealt{Genel2019, Keller2019}). The reduction in the final stellar mass (legend) is therefore primarily due to the dearth of old stars.

We therefore conclude that uncertainties in the timing of hydrogen reionization primarily impact the proportion of old to young stars in low-mass dwarfs, rather than their ability to reignite star formation at late times. Similar inhomogeneous heating is expected from the reionization of helium by quasars at later times. This phase transition is expected to be complete by $z \sim 2$ at the latest (e.g. \citealt{McQuinn2009}) and is therefore unlikely to create an environmental modulation of the reignition of star formation at $z \leq 1$. The overall level of heating due to helium reionization remains uncertain (e.g. \citealt{UptonSanderbeck2016}) and we defer to future work a characterization of the impact of helium reionization on our results. Finally, we note that recent observations indicate that the \hi \ photoionization rates had been previously understimated at low redshift ($z \leq 0.5$; \citealt{Gaikwad2017, Khaire2019}), potentially modifying the detailed balance of heating and cooling at late times. These new constraints have just been incorporated in the latest iterations of uniform UV background models (\citealt{Puchwein2019, FaucherGiguere2020}), and we leave a more detailed exploration of their importance to future work.

\subsubsection{Galaxy formation physics} \label{sec:results:physics:gf}

\begin{figure}
  \centering
    \includegraphics[width=\columnwidth]{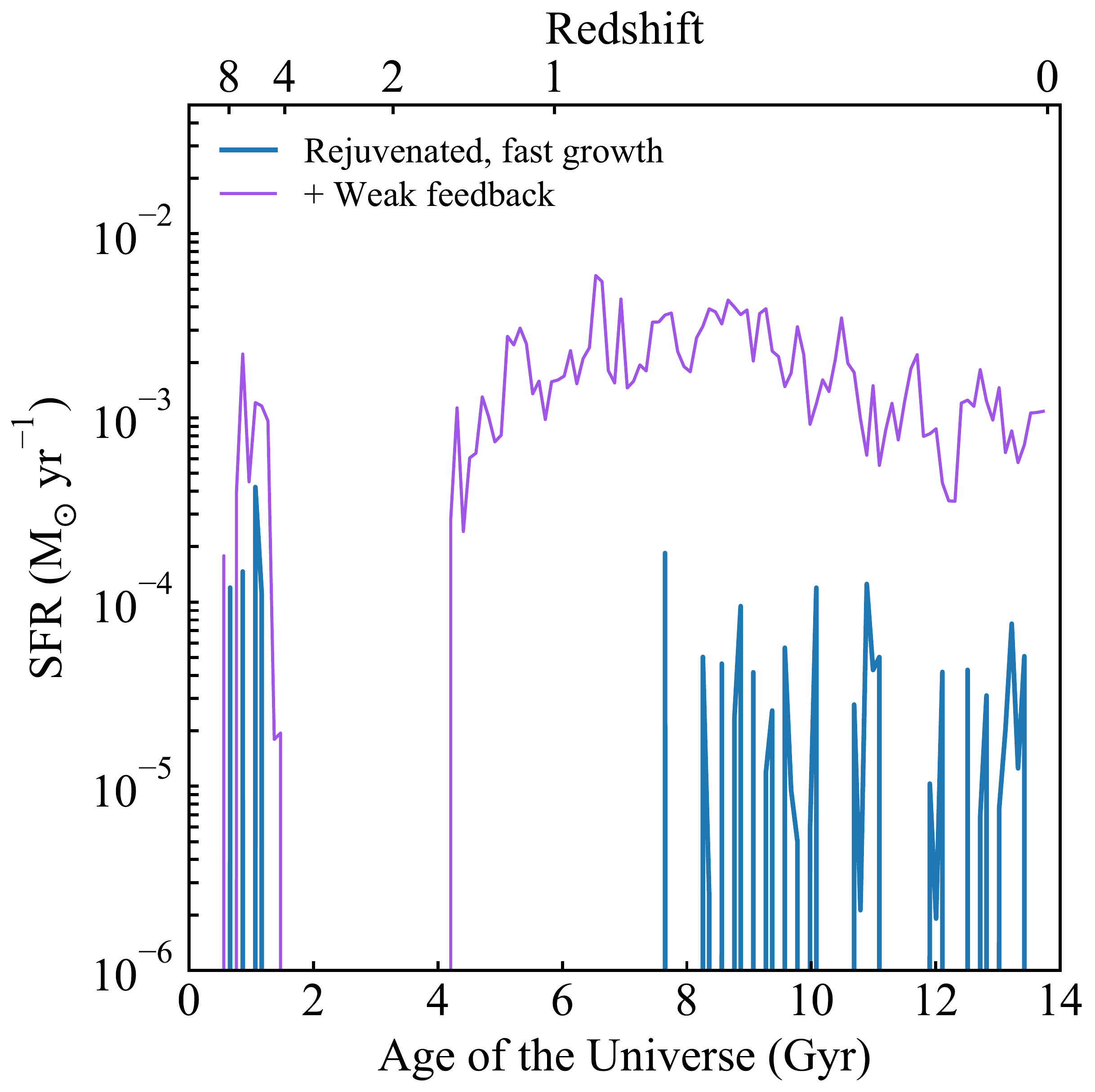}

    \caption{Response of star formation to large variations in our feedback modelling. Unphysically weakening the feedback efficiency (violet) quickens rejuvenation compared to our fiducial model (blue), as more gas is retained within the halo after reionization and residual feedback from SNe Ia and AGB winds is less effective.}
    \label{fig:physics}
\end{figure}

We now probe whether our results are robust to a large variation in our feedback implementation. We re-evolve the same dwarf galaxy (`Rejuvenated, fast growth') to $z=0$ with an alternative, `Weak feedback' model. This model introduces temperature and velocity ceilings for SN and wind ejectas, thereby limiting their efficiency in regulating star formation (see \citealt{Agertz2020} for more details). Figure~\ref{fig:physics} compares the star formation histories of the fiducial (blue) and the `Weak feedback' models (violet). The reignition of star formation at late times is present in both models, although weaker feedback permits higher star formation rates and a higher final stellar mass (see Table~\ref{table:runs}). Reignition is faster with weaker feedback by several billion years -- the reason is twofold: (i) With less ejective feedback from massive stars, more gas remains within the central kiloparsec of the halo at $z=4$ ($\xMsol{4}{5}$ compared to \xMsol{0.7}{5} in the fiducial case), in turn accelerating the gas accretion phase; and (ii) less efficient feedback from old stellar populations at late times permits faster rejuvenation in agreement with Section~\ref{sec:results:residualFB}. The `Weak feedback' model is not a physically motivated prescription, overproducing stars by nature -- we use it as a demonstration that feedback from both old stellar populations and young, massive stars impacts the later ability of a dwarf to reignite star formation. Despite being active several billion years before reignition, massive stars define the state and amount of gas in the halo shortly after reionization, i.e. setting the initial conditions for the transition towards rejuvenation. 

We next explore the importance of numerical resolution in our results. We re-evolve the same galaxy using our fiducial model, but increasing the mass resolution from $m_{\text{DM}} = 960$ to $120 \, \Msol$, while keeping the same maximum spatial refinement of $3 \, \pc$. We find that increasing the mass resolution nearly doubles the final stellar mass of the galaxy to $\xMsol{9}{5}$, consistent with previous findings in \citet{Agertz2020}. The reignition of star formation at late times is also observed at higher resolution, at a nearly identical time ($z=0.61$ compared to $z=0.64$ in the fiducial case) and average rate ($\xScientific{1.7}{-5} \, \Msolyr$ compared to $\xScientific{1.5}{-5} \, \Msolyr$). We conclude that, despite uncertainties on the final stellar masses of our galaxies due to the coupling between feedback and numerical resolution, the existence and timing of rejuvenation is converged with resolution.

We have established that the reignition of star formation in a dwarf galaxy depends on the balance between cooling, mass growth (Section~\ref{sec:results:gms}) and heating from reionization and stars (Sections~\ref{sec:results:residualFB} and~\ref{sec:results:physics:reionization}). Other inputs to this equilibrium, however, remain uncertain, and could add to the variety of physical processes highlighted in this work. Notably, future studies that wish to determine the absolute boundary of star formation in field dwarf galaxies will need to investigate how to account for non-equilibrium and molecular cooling (e.g. \citealt{Gnedin2011, Christensen2012, Nickerson2018}) and how to model the formation of stars at such high resolutions (e.g. \citealt{Su2018, Munshi2019, Applebaum2020}). This sensitivity will be key to improving constraints on star formation, feedback, and reionization with the next generation of observations of dwarf galaxies.

\section{Conclusions} \label{sec:conclusion}

We have presented results showing how low-mass, field dwarf galaxies ($10^5 \lesssim \Mstar \lesssim 10^6 \, \Msol$) can reignite and sustain star formation at late times. We presented a suite of four galaxies with increasing halo mass today, evolved to $z=0$ using high-resolution cosmological simulations (\citealt{Rey2019b, Agertz2020}). We complemented this main suite with experiments varying either (i) the implementation of feedback and (ii) the mass build-up of galaxies, using the genetic modification approach (\citealt{Roth2016, Rey2018}). 

All dwarf galaxies in our suite have sufficiently low masses to be quenched by the reduction of gas inflows once cosmic reionization has heated their surrounding medium. However, the two more massive galaxies ($\Mvir \approx \xMsol{3}{9}$ at $z=0$) both experience a reignition of star formation after $z=1$, forming a significant fraction of their stellar mass \textit{in situ} at low average star formation rates ($10^{-5} \, \Msolyr$; see Figure~\ref{fig:massgrowth}). 

We link the rejuvenation of star formation to a slow accretion of gas which steadily increases the central gas mass of the dwarf until reignition (Figure~\ref{fig:gasmass}). After reionization, growth in dynamical mass allows leftover gas within the halo to reach sufficient densities to cool efficiently. Such gas then starts condensing in a near-isothermal equilibrium between radiative cooling and photoheating from the surrounding UV background. Once self-shielding densities are reached, the gas decouples from the UV background and is allowed to cool below $10^4\,$K to fuel star formation (Figure~\ref{fig:gasphase}). 

Because of the low gas masses within these dwarfs, we find that small energy inputs can balance the cooling of gas and delay the onset of star formation. We demonstrate that the combination of SNe Ia and AGB stars, which are active long after quenching by reionization, are sufficient to lengthen the quiescent period by several billion years (Figure~\ref{fig:agbs}). 

The time necessary to complete the rejuvenation of star formation also strongly interacts with the specific mass growth history of each dwarf galaxy. Using the genetic modification framework, we demonstrate that galaxies in which rejuvenation is prevented by residual feedback can overcome this barrier by a sufficient increase in dynamical mass at late times. Furthermore, we show that slowing the mass build-up of a rejuvenating galaxy delays, and nearly prevents, the reignition of star formation at fixed halo mass today (Figure~\ref{fig:gms}).

Our results extend previous studies of the modulation of star formation histories by the UV background in dwarf galaxies (e.g. \citealt{BenitezLLambay2015, Fitts2017, Jeon2017}). We reaffirm that the competition between the depth of the potential well and the temperature of the IGM is the primary driver for the reignition of star formation at late times (e.g. \citealt{Efstathiou1992, Gnedin2000, Hoeft2006, Okamoto2008, Noh2014}). However, our results demonstrate that galaxy formation physics, and in particular feedback, adds to this interplay and plays a key role in dictating the ability of a dwarf galaxy to rejuvenate. In addition to the sensitivity to feedback from old stellar populations (Figure~\ref{fig:agbs}), we show that the nature of feedback from massive stars, which regulate the first star-forming phase and hence the state of the galaxy after reionization has ended, also matters (Figure~\ref{fig:physics}). Uncertainties in modelling galaxy formation physics -- in particular cooling, heating, and feedback -- are likely to account for discrepancies in the reported, minimum dynamical masses to sustain star formation in isolated dwarfs, ranging from $\Mvir \simeq \xScientific{3}{9}$ (this work; \citealt{Jeon2017}) to $\Mvir \gtrsim \xMsol{7}{9}$ (e.g. \citealt{BenitezLLambay2015, Fitts2017, Ledinauskas2018, Wheeler2019}).

Despite these uncertainties, the low levels of star formation seen in our simulated galaxies highly resembles those observed in, for example, Leo~T and Leo~P (\citealt{Clementini2012, Weisz2012, McQuinn2015LeoP}), the smallest star-forming dwarfs currently known. We will produce a detailed observational comparison with our simulated galaxies in a follow-up work, focussing on the star formation histories and gas kinematics. Furthermore, mechanisms uncovered in this work are likely generic across dwarf galaxies, and naturally predict a diversity in stellar populations and gas content at the low-mass end. We have shown examples of (i) gas-rich, low-mass dwarfs with ongoing star formation, (ii) low-mass dwarfs caught in transition, exhibiting an exclusively old stellar population but a sizeable gas content, and (iii) gas-poor, quenched low-mass dwarfs. All three can be hosted by dark matter haloes of a similar mass. Finding these different classes of galaxies is within the reach of future experiments, as the combination of current-generation deep imaging with \hi \ surveys has already resulted in possible detections of gas-rich, quenched faint dwarfs (\citealt{Janesh2019}). Deeper imaging, with e.g. LSST, will directly probe the formation of \MR{faint field} dwarfs, highlighting the need for predictions of the abundances and detectabilities of such systems in the Local Volume. We will tackle this requirement in future work, using a semi-empirical model based on these results to obtain a larger sample of galaxies (Kim et al. in preparation). 

\section*{Acknowledgements}
MR thanks Eric Andersson, Alyson Brooks, George Efstathiou, Richard Ellis, Pierre Ocvirk, Florent Renaud, Am\'elie Saintonge, Romain Teyssier, and Anna Wright for helpful discussions during the construction of this work. \MR{The authors would like to thank the anonymous referee for a constructive report that improved the quality of this paper}. MR acknowledges support from the Perren Fund, the IMPACT Fund, and the Knut and Alice Wallenberg Foundation. AP is supported by the Royal Society. OA acknowledges support from the Swedish Research Council (grant 2014-5791) and the Knut and Alice Wallenberg Foundation. MO would like to thank the STFC for support (grant ST/R505134/1). This project has received funding from the European Union Horizon 2020 research and innovation programme under grant agreement No. 818085 GMGalaxies. This work was partially supported by the UCL Cosmoparticle Initiative. The authors acknowledge the use of the UCL Grace High Performance Computing Facility, the Surrey Eureka supercomputer facility, and associated support services. This work was performed in part using the DiRAC Data Intensive service at Leicester, operated by the University of Leicester IT Services, which is part of the STFC DiRAC HPC Facility (www.dirac.ac.uk).




\bibliographystyle{mnras}
\bibliography{Biblio} 

\begin{thebibliography}{}
\makeatletter
\relax
\def\mn@urlcharsother{\let\do\@makeother \do\$\do\&\do\#\do\^\do\_\do\%\do\~}
\def\mn@doi{\begingroup\mn@urlcharsother \@ifnextchar [ {\mn@doi@}
  {\mn@doi@[]}}
\def\mn@doi@[#1]#2{\def\@tempa{#1}\ifx\@tempa\@empty \href
  {http://dx.doi.org/#2} {doi:#2}\else \href {http://dx.doi.org/#2} {#1}\fi
  \endgroup}
\def\mn@eprint#1#2{\mn@eprint@#1:#2::\@nil}
\def\mn@eprint@arXiv#1{\href {http://arxiv.org/abs/#1} {{\tt arXiv:#1}}}
\def\mn@eprint@dblp#1{\href {http://dblp.uni-trier.de/rec/bibtex/#1.xml}
  {dblp:#1}}
\def\mn@eprint@#1:#2:#3:#4\@nil{\def\@tempa {#1}\def\@tempb {#2}\def\@tempc
  {#3}\ifx \@tempc \@empty \let \@tempc \@tempb \let \@tempb \@tempa \fi \ifx
  \@tempb \@empty \def\@tempb {arXiv}\fi \@ifundefined
  {mn@eprint@\@tempb}{\@tempb:\@tempc}{\expandafter \expandafter \csname
  mn@eprint@\@tempb\endcsname \expandafter{\@tempc}}}

\bibitem[\protect\citeauthoryear{{Adams} \& {Oosterloo}}{{Adams} \&
  {Oosterloo}}{2018}]{Adams2018}
{Adams} E. A.~K.,  {Oosterloo} T.~A.,  2018, \mn@doi [\aap]
  {10.1051/0004-6361/201732017}, 612, A26

\bibitem[\protect\citeauthoryear{{Agertz}, {Teyssier}  \& {Moore}}{{Agertz}
  et~al.}{2011}]{Agertz2011}
{Agertz} O.,  {Teyssier} R.,   {Moore} B.,  2011, \mn@doi [\mnras]
  {10.1111/j.1365-2966.2010.17530.x}, 410, 1391

\bibitem[\protect\citeauthoryear{{Agertz}, {Kravtsov}, {Leitner}  \&
  {Gnedin}}{{Agertz} et~al.}{2013}]{Agertz2013}
{Agertz} O.,  {Kravtsov} A.~V.,  {Leitner} S.~N.,   {Gnedin} N.~Y.,  2013,
  \mn@doi [\apj] {10.1088/0004-637X/770/1/25}, 770, 25

\bibitem[\protect\citeauthoryear{{Agertz} et~al.,}{{Agertz}
  et~al.}{2020}]{Agertz2020}
{Agertz} O.,  et~al., 2020, \mn@doi [\mnras] {10.1093/mnras/stz3053}, 491, 1656

\bibitem[\protect\citeauthoryear{{Applebaum}, {Brooks}, {Quinn}  \&
  {Christensen}}{{Applebaum} et~al.}{2020}]{Applebaum2020}
{Applebaum} E.,  {Brooks} A.~M.,  {Quinn} T.~R.,   {Christensen} C.~R.,  2020,
  \mn@doi [\mnras] {10.1093/mnras/stz3331}, 492, 8

\bibitem[\protect\citeauthoryear{{Aubert} \& {Teyssier}}{{Aubert} \&
  {Teyssier}}{2010}]{Aubert2010}
{Aubert} D.,  {Teyssier} R.,  2010, \mn@doi [\apj]
  {10.1088/0004-637X/724/1/244}, 724, 244

\bibitem[\protect\citeauthoryear{{Aubert} et~al.,}{{Aubert}
  et~al.}{2018}]{Aubert2018}
{Aubert} D.,  et~al., 2018, \mn@doi [\apjl] {10.3847/2041-8213/aab14d}, 856,
  L22

\bibitem[\protect\citeauthoryear{{Ben{\'\i}tez-Llambay}, {Navarro}, {Abadi},
  {Gottl{\"o}ber}, {Yepes}, {Hoffman}  \& {Steinmetz}}{{Ben{\'\i}tez-Llambay}
  et~al.}{2015}]{BenitezLLambay2015}
{Ben{\'\i}tez-Llambay} A.,  {Navarro} J.~F.,  {Abadi} M.~G.,  {Gottl{\"o}ber}
  S.,  {Yepes} G.,  {Hoffman} Y.,   {Steinmetz} M.,  2015, \mn@doi [\mnras]
  {10.1093/mnras/stv925}, 450, 4207

\bibitem[\protect\citeauthoryear{{Benson}, {Lacey}, {Baugh}, {Cole}  \&
  {Frenk}}{{Benson} et~al.}{2002}]{Benson2002}
{Benson} A.~J.,  {Lacey} C.~G.,  {Baugh} C.~M.,  {Cole} S.,   {Frenk} C.~S.,
  2002, \mn@doi [\mnras] {10.1046/j.1365-8711.2002.05387.x}, 333, 156

\bibitem[\protect\citeauthoryear{{Bolton}, {Becker}, {Haehnelt}  \&
  {Viel}}{{Bolton} et~al.}{2014}]{Bolton2014}
{Bolton} J.~S.,  {Becker} G.~D.,  {Haehnelt} M.~G.,   {Viel} M.,  2014, \mn@doi
  [\mnras] {10.1093/mnras/stt2374}, 438, 2499

\bibitem[\protect\citeauthoryear{{Brown} et~al.,}{{Brown}
  et~al.}{2014}]{Brown2014}
{Brown} T.~M.,  et~al., 2014, \mn@doi [\apj] {10.1088/0004-637X/796/2/91}, 796,
  91

\bibitem[\protect\citeauthoryear{{Brunker} et~al.,}{{Brunker}
  et~al.}{2019}]{Brunker2019}
{Brunker} S.~W.,  et~al., 2019, \mn@doi [\aj] {10.3847/1538-3881/aafb39}, 157,
  76

\bibitem[\protect\citeauthoryear{{Bullock}, {Kravtsov}  \&
  {Weinberg}}{{Bullock} et~al.}{2000}]{Bullock2000}
{Bullock} J.~S.,  {Kravtsov} A.~V.,   {Weinberg} D.~H.,  2000, \mn@doi [\apj]
  {10.1086/309279}, 539, 517

\bibitem[\protect\citeauthoryear{{Christensen}, {Quinn}, {Governato}, {Stilp},
  {Shen}  \& {Wadsley}}{{Christensen} et~al.}{2012}]{Christensen2012}
{Christensen} C.,  {Quinn} T.,  {Governato} F.,  {Stilp} A.,  {Shen} S.,
  {Wadsley} J.,  2012, \mn@doi [\mnras] {10.1111/j.1365-2966.2012.21628.x},
  \href {https://ui.adsabs.harvard.edu/abs/2012MNRAS.425.3058C} {425, 3058}

\bibitem[\protect\citeauthoryear{{Clementini}, {Cignoni}, {Contreras Ramos},
  {Federici}, {Ripepi}, {Marconi}, {Tosi}  \& {Musella}}{{Clementini}
  et~al.}{2012}]{Clementini2012}
{Clementini} G.,  {Cignoni} M.,  {Contreras Ramos} R.,  {Federici} L.,
  {Ripepi} V.,  {Marconi} M.,  {Tosi} M.,   {Musella} I.,  2012, \mn@doi [\apj]
  {10.1088/0004-637X/756/2/108}, 756, 108

\bibitem[\protect\citeauthoryear{{Compostella}, {Cantalupo}  \&
  {Porciani}}{{Compostella} et~al.}{2013}]{Compostella2013}
{Compostella} M.,  {Cantalupo} S.,   {Porciani} C.,  2013, \mn@doi [\mnras]
  {10.1093/mnras/stt1510}, 435, 3169

\bibitem[\protect\citeauthoryear{{Conroy}, {van Dokkum}  \&
  {Kravtsov}}{{Conroy} et~al.}{2015}]{Conroy2015}
{Conroy} C.,  {van Dokkum} P.~G.,   {Kravtsov} A.,  2015, \mn@doi [\apj]
  {10.1088/0004-637X/803/2/77}, 803, 77

\bibitem[\protect\citeauthoryear{{Courty} \& {Alimi}}{{Courty} \&
  {Alimi}}{2004}]{Courty2004}
{Courty} S.,  {Alimi} J.~M.,  2004, \mn@doi [\aap]
  {10.1051/0004-6361:20031736}, 416, 875

\bibitem[\protect\citeauthoryear{{DeFelippis}, {Putman}  \&
  {Tollerud}}{{DeFelippis} et~al.}{2019}]{DeFelippis2019}
{DeFelippis} D.,  {Putman} M.,   {Tollerud} E.,  2019, \mn@doi [\apj]
  {10.3847/1538-4357/ab1e57}, 879, 22

\bibitem[\protect\citeauthoryear{{Di Cintio}, {Brook}, {Macci{\`o}}, {Stinson},
  {Knebe}, {Dutton}  \& {Wadsley}}{{Di Cintio} et~al.}{2014}]{DiCintio2014}
{Di Cintio} A.,  {Brook} C.~B.,  {Macci{\`o}} A.~V.,  {Stinson} G.~S.,  {Knebe}
  A.,  {Dutton} A.~A.,   {Wadsley} J.,  2014, \mn@doi [\mnras]
  {10.1093/mnras/stt1891}, 437, 415

\bibitem[\protect\citeauthoryear{{Efstathiou}}{{Efstathiou}}{1992}]{Efstathiou1992}
{Efstathiou} G.,  1992, \mn@doi [\mnras] {10.1093/mnras/256.1.43P}, 256, 43P

\bibitem[\protect\citeauthoryear{{Eisenstein} \& {Hut}}{{Eisenstein} \&
  {Hut}}{1998}]{Eisenstein1998}
{Eisenstein} D.~J.,  {Hut} P.,  1998, \mn@doi [\apj] {10.1086/305535}, 498, 137

\bibitem[\protect\citeauthoryear{{Errani}, {Pe{\~n}arrubia}  \&
  {Walker}}{{Errani} et~al.}{2018}]{Errani2018}
{Errani} R.,  {Pe{\~n}arrubia} J.,   {Walker} M.~G.,  2018, \mn@doi [\mnras]
  {10.1093/mnras/sty2505}, 481, 5073

\bibitem[\protect\citeauthoryear{{Faucher-Gigu{\`e}re}}{{Faucher-Gigu{\`e}re}}{2020}]{FaucherGiguere2020}
{Faucher-Gigu{\`e}re} C.-A.,  2020, \mn@doi [\mnras] {10.1093/mnras/staa302},
  493, 1614

\bibitem[\protect\citeauthoryear{{Faucher-Gigu{\`e}re}, {Kere{\v{s}}},
  {Dijkstra}, {Hernquist}  \& {Zaldarriaga}}{{Faucher-Gigu{\`e}re}
  et~al.}{2010}]{FaucherGiguere2010}
{Faucher-Gigu{\`e}re} C.-A.,  {Kere{\v{s}}} D.,  {Dijkstra} M.,  {Hernquist}
  L.,   {Zaldarriaga} M.,  2010, \mn@doi [\apj] {10.1088/0004-637X/725/1/633},
  \href {https://ui.adsabs.harvard.edu/abs/2010ApJ...725..633F} {725, 633}

\bibitem[\protect\citeauthoryear{{Ferland}, {Korista}, {Verner}, {Ferguson},
  {Kingdon}  \& {Verner}}{{Ferland} et~al.}{1998}]{Ferland1998}
{Ferland} G.~J.,  {Korista} K.~T.,  {Verner} D.~A.,  {Ferguson} J.~W.,
  {Kingdon} J.~B.,   {Verner} E.~M.,  1998, \mn@doi [\pasp] {10.1086/316190},
  110, 761

\bibitem[\protect\citeauthoryear{{Finlator}, {Keating}, {Oppenheimer},
  {Dav{\'e}}  \& {Zackrisson}}{{Finlator} et~al.}{2018}]{Finlator2018}
{Finlator} K.,  {Keating} L.,  {Oppenheimer} B.~D.,  {Dav{\'e}} R.,
  {Zackrisson} E.,  2018, \mn@doi [\mnras] {10.1093/mnras/sty1949}, 480, 2628

\bibitem[\protect\citeauthoryear{{Fitts} et~al.,}{{Fitts}
  et~al.}{2017}]{Fitts2017}
{Fitts} A.,  et~al., 2017, \mn@doi [\mnras] {10.1093/mnras/stx1757}, 471, 3547

\bibitem[\protect\citeauthoryear{{Forbes}, {Read}, {Gieles}  \&
  {Collins}}{{Forbes} et~al.}{2018}]{Forbes2018}
{Forbes} D.~A.,  {Read} J.~I.,  {Gieles} M.,   {Collins} M. L.~M.,  2018,
  \mn@doi [\mnras] {10.1093/mnras/sty2584}, 481, 5592

\bibitem[\protect\citeauthoryear{{Gaikwad}, {Khaire}, {Choudhury}  \&
  {Srianand}}{{Gaikwad} et~al.}{2017}]{Gaikwad2017}
{Gaikwad} P.,  {Khaire} V.,  {Choudhury} T.~R.,   {Srianand} R.,  2017, \mn@doi
  [\mnras] {10.1093/mnras/stw3086}, 466, 838

\bibitem[\protect\citeauthoryear{{Genel} et~al.,}{{Genel}
  et~al.}{2019}]{Genel2019}
{Genel} S.,  et~al., 2019, \mn@doi [\apj] {10.3847/1538-4357/aaf4bb}, 871, 21

\bibitem[\protect\citeauthoryear{{Girardi} et~al.,}{{Girardi}
  et~al.}{2010}]{Girardi2010}
{Girardi} L.,  et~al., 2010, \mn@doi [\apj] {10.1088/0004-637X/724/2/1030},
  724, 1030

\bibitem[\protect\citeauthoryear{{Gnedin}}{{Gnedin}}{2000}]{Gnedin2000}
{Gnedin} N.~Y.,  2000, \mn@doi [\apj] {10.1086/317042}, 542, 535

\bibitem[\protect\citeauthoryear{{Gnedin} \& {Kravtsov}}{{Gnedin} \&
  {Kravtsov}}{2011}]{Gnedin2011}
{Gnedin} N.~Y.,  {Kravtsov} A.~V.,  2011, \mn@doi [\apj]
  {10.1088/0004-637X/728/2/88}, 728, 88

\bibitem[\protect\citeauthoryear{{Guillet} \& {Teyssier}}{{Guillet} \&
  {Teyssier}}{2011}]{Guillet2011}
{Guillet} T.,  {Teyssier} R.,  2011, \mn@doi [Journal of Computational Physics]
  {10.1016/j.jcp.2011.02.044}, 230, 4756

\bibitem[\protect\citeauthoryear{{Haardt} \& {Madau}}{{Haardt} \&
  {Madau}}{1996}]{Haardt1996}
{Haardt} F.,  {Madau} P.,  1996, \mn@doi [\apj] {10.1086/177035}, 461, 20

\bibitem[\protect\citeauthoryear{{Hargis} et~al.,}{{Hargis}
  et~al.}{2020}]{Hargis2020}
{Hargis} J.~R.,  et~al., 2020, \mn@doi [\apj] {10.3847/1538-4357/ab58d2}, 888,
  31

\bibitem[\protect\citeauthoryear{{Haynes} et~al.,}{{Haynes}
  et~al.}{2011}]{Haynes2011}
{Haynes} M.~P.,  et~al., 2011, \mn@doi [\aj] {10.1088/0004-6256/142/5/170},
  142, 170

\bibitem[\protect\citeauthoryear{{Hiss}, {Walther}, {Hennawi}, {O{\~n}orbe},
  {O'Meara}, {Rorai}  \& {Luki{\'c}}}{{Hiss} et~al.}{2018}]{Hiss2018}
{Hiss} H.,  {Walther} M.,  {Hennawi} J.~F.,  {O{\~n}orbe} J.,  {O'Meara} J.~M.,
   {Rorai} A.,   {Luki{\'c}} Z.,  2018, \mn@doi [\apj]
  {10.3847/1538-4357/aada86}, 865, 42

\bibitem[\protect\citeauthoryear{{Hoeft}, {Yepes}, {Gottl{\"o}ber}  \&
  {Springel}}{{Hoeft} et~al.}{2006}]{Hoeft2006}
{Hoeft} M.,  {Yepes} G.,  {Gottl{\"o}ber} S.,   {Springel} V.,  2006, \mn@doi
  [\mnras] {10.1111/j.1365-2966.2006.10678.x}, 371, 401

\bibitem[\protect\citeauthoryear{{Iliev}, {Mellema}, {Ahn}, {Shapiro}, {Mao}
  \& {Pen}}{{Iliev} et~al.}{2014}]{Iliev2014}
{Iliev} I.~T.,  {Mellema} G.,  {Ahn} K.,  {Shapiro} P.~R.,  {Mao} Y.,   {Pen}
  U.-L.,  2014, \mn@doi [\mnras] {10.1093/mnras/stt2497}, 439, 725

\bibitem[\protect\citeauthoryear{{Irwin} et~al.,}{{Irwin}
  et~al.}{2007}]{Irwin2007}
{Irwin} M.~J.,  et~al., 2007, \mn@doi [\apjl] {10.1086/512183}, 656, L13

\bibitem[\protect\citeauthoryear{{Janesh}, {Rhode}, {Salzer}, {Janowiecki},
  {Adams}, {Haynes}, {Giovanelli}  \& {Cannon}}{{Janesh}
  et~al.}{2019}]{Janesh2019}
{Janesh} W.,  {Rhode} K.~L.,  {Salzer} J.~J.,  {Janowiecki} S.,  {Adams} E.
  A.~K.,  {Haynes} M.~P.,  {Giovanelli} R.,   {Cannon} J.~M.,  2019, \mn@doi
  [\aj] {10.3847/1538-3881/ab12d3}, 157, 183

\bibitem[\protect\citeauthoryear{{Jeon}, {Besla}  \& {Bromm}}{{Jeon}
  et~al.}{2017}]{Jeon2017}
{Jeon} M.,  {Besla} G.,   {Bromm} V.,  2017, \mn@doi [\apj]
  {10.3847/1538-4357/aa8c80}, 848, 85

\bibitem[\protect\citeauthoryear{{Jethwa}, {Erkal}  \& {Belokurov}}{{Jethwa}
  et~al.}{2018}]{Jethwa2018}
{Jethwa} P.,  {Erkal} D.,   {Belokurov} V.,  2018, \mn@doi [\mnras]
  {10.1093/mnras/stx2330}, 473, 2060

\bibitem[\protect\citeauthoryear{{Katz} et~al.,}{{Katz}
  et~al.}{2019}]{Katz2019}
{Katz} H.,  et~al., 2019, arXiv e-prints, p. arXiv:1905.11414

\bibitem[\protect\citeauthoryear{{Keating}, {Weinberger}, {Kulkarni},
  {Haehnelt}, {Chardin}  \& {Aubert}}{{Keating} et~al.}{2020}]{Keating2020}
{Keating} L.~C.,  {Weinberger} L.~H.,  {Kulkarni} G.,  {Haehnelt} M.~G.,
  {Chardin} J.,   {Aubert} D.,  2020, \mn@doi [\mnras] {10.1093/mnras/stz3083},
  491, 1736

\bibitem[\protect\citeauthoryear{{Keller}, {Wadsley}, {Wang}  \&
  {Kruijssen}}{{Keller} et~al.}{2019}]{Keller2019}
{Keller} B.~W.,  {Wadsley} J.~W.,  {Wang} L.,   {Kruijssen} J.~M.~D.,  2019,
  \mn@doi [\mnras] {10.1093/mnras/sty2859}, 482, 2244

\bibitem[\protect\citeauthoryear{{Khaire} et~al.,}{{Khaire}
  et~al.}{2019}]{Khaire2019}
{Khaire} V.,  et~al., 2019, \mn@doi [\mnras] {10.1093/mnras/stz344}, 486, 769

\bibitem[\protect\citeauthoryear{{Kim} \& {Ostriker}}{{Kim} \&
  {Ostriker}}{2015}]{Kim2015}
{Kim} C.-G.,  {Ostriker} E.~C.,  2015, \mn@doi [\apj]
  {10.1088/0004-637X/815/1/67}, 815, 67

\bibitem[\protect\citeauthoryear{{Kroupa}}{{Kroupa}}{2001}]{Kroupa2001}
{Kroupa} P.,  2001, \mn@doi [\mnras] {10.1046/j.1365-8711.2001.04022.x}, 322,
  231

\bibitem[\protect\citeauthoryear{{Ledinauskas} \& {Zubovas}}{{Ledinauskas} \&
  {Zubovas}}{2018}]{Ledinauskas2018}
{Ledinauskas} E.,  {Zubovas} K.,  2018, \mn@doi [\aap]
  {10.1051/0004-6361/201832824}, 615, A64

\bibitem[\protect\citeauthoryear{{Macci{\`o}}, {Frings}, {Buck}, {Penzo},
  {Dutton}, {Blank}  \& {Obreja}}{{Macci{\`o}} et~al.}{2017}]{Maccio2017}
{Macci{\`o}} A.~V.,  {Frings} J.,  {Buck} T.,  {Penzo} C.,  {Dutton} A.~A.,
  {Blank} M.,   {Obreja} A.,  2017, \mn@doi [\mnras] {10.1093/mnras/stx2048},
  472, 2356

\bibitem[\protect\citeauthoryear{{Mannucci}, {Della Valle}  \&
  {Panagia}}{{Mannucci} et~al.}{2006}]{Mannucci2006}
{Mannucci} F.,  {Della Valle} M.,   {Panagia} N.,  2006, \mn@doi [\mnras]
  {10.1111/j.1365-2966.2006.10501.x}, 370, 773

\bibitem[\protect\citeauthoryear{{Maoz}, {Mannucci}  \& {Brandt}}{{Maoz}
  et~al.}{2012}]{Maoz2012}
{Maoz} D.,  {Mannucci} F.,   {Brandt} T.~D.,  2012, \mn@doi [\mnras]
  {10.1111/j.1365-2966.2012.21871.x}, 426, 3282

\bibitem[\protect\citeauthoryear{{Martizzi}, {Faucher-Gigu{\`e}re}  \&
  {Quataert}}{{Martizzi} et~al.}{2015}]{Martizzi2015}
{Martizzi} D.,  {Faucher-Gigu{\`e}re} C.-A.,   {Quataert} E.,  2015, \mn@doi
  [\mnras] {10.1093/mnras/stv562}, 450, 504

\bibitem[\protect\citeauthoryear{{McQuinn}}{{McQuinn}}{2016}]{McQuinn2016_review}
{McQuinn} M.,  2016, \mn@doi [\araa] {10.1146/annurev-astro-082214-122355}, 54,
  313

\bibitem[\protect\citeauthoryear{{McQuinn} \& {Upton Sanderbeck}}{{McQuinn} \&
  {Upton Sanderbeck}}{2016}]{McQuinn2016_TIGM}
{McQuinn} M.,  {Upton Sanderbeck} P.~R.,  2016, \mn@doi [\mnras]
  {10.1093/mnras/stv2675}, 456, 47

\bibitem[\protect\citeauthoryear{{McQuinn}, {Lidz}, {Zaldarriaga}, {Hernquist},
  {Hopkins}, {Dutta}  \& {Faucher-Gigu{\`e}re}}{{McQuinn}
  et~al.}{2009}]{McQuinn2009}
{McQuinn} M.,  {Lidz} A.,  {Zaldarriaga} M.,  {Hernquist} L.,  {Hopkins} P.~F.,
   {Dutta} S.,   {Faucher-Gigu{\`e}re} C.-A.,  2009, \mn@doi [\apj]
  {10.1088/0004-637X/694/2/842}, 694, 842

\bibitem[\protect\citeauthoryear{{McQuinn} et~al.,}{{McQuinn}
  et~al.}{2015}]{McQuinn2015LeoP}
{McQuinn} K. B.~W.,  et~al., 2015, \mn@doi [\apj]
  {10.1088/0004-637X/812/2/158}, 812, 158

\bibitem[\protect\citeauthoryear{{McQuinn} et~al.,}{{McQuinn}
  et~al.}{2020}]{McQuinn2020}
{McQuinn} K. B.~W.,  et~al., 2020, \mn@doi [\apj] {10.3847/1538-4357/ab7447},
  891, 181

\bibitem[\protect\citeauthoryear{{Munshi}, {Brooks}, {Christensen},
  {Applebaum}, {Holley-Bockelmann}, {Quinn}  \& {Wadsley}}{{Munshi}
  et~al.}{2019}]{Munshi2019}
{Munshi} F.,  {Brooks} A.~M.,  {Christensen} C.,  {Applebaum} E.,
  {Holley-Bockelmann} K.,  {Quinn} T.~R.,   {Wadsley} J.,  2019, \mn@doi [\apj]
  {10.3847/1538-4357/ab0085}, 874, 40

\bibitem[\protect\citeauthoryear{{Nickerson}, {Teyssier}  \&
  {Rosdahl}}{{Nickerson} et~al.}{2018}]{Nickerson2018}
{Nickerson} S.,  {Teyssier} R.,   {Rosdahl} J.,  2018, \mn@doi [\mnras]
  {10.1093/mnras/sty1556}, 479, 3206

\bibitem[\protect\citeauthoryear{{Noh} \& {McQuinn}}{{Noh} \&
  {McQuinn}}{2014}]{Noh2014}
{Noh} Y.,  {McQuinn} M.,  2014, \mn@doi [\mnras] {10.1093/mnras/stu1412}, 444,
  503

\bibitem[\protect\citeauthoryear{{O{\~n}orbe}, {Boylan-Kolchin}, {Bullock},
  {Hopkins}, {Kere{\v{s}}}, {Faucher-Gigu{\`e}re}, {Quataert}  \&
  {Murray}}{{O{\~n}orbe} et~al.}{2015}]{Onorbe2015}
{O{\~n}orbe} J.,  {Boylan-Kolchin} M.,  {Bullock} J.~S.,  {Hopkins} P.~F.,
  {Kere{\v{s}}} D.,  {Faucher-Gigu{\`e}re} C.-A.,  {Quataert} E.,   {Murray}
  N.,  2015, \mn@doi [\mnras] {10.1093/mnras/stv2072}, 454, 2092

\bibitem[\protect\citeauthoryear{{O{\~n}orbe}, {Hennawi}  \&
  {Luki{\'c}}}{{O{\~n}orbe} et~al.}{2017}]{Onorbe2017}
{O{\~n}orbe} J.,  {Hennawi} J.~F.,   {Luki{\'c}} Z.,  2017, \mn@doi [\apj]
  {10.3847/1538-4357/aa6031}, 837, 106

\bibitem[\protect\citeauthoryear{{O'Shea}, {Wise}, {Xu}  \& {Norman}}{{O'Shea}
  et~al.}{2015}]{OShea2015}
{O'Shea} B.~W.,  {Wise} J.~H.,  {Xu} H.,   {Norman} M.~L.,  2015, \mn@doi
  [\apjl] {10.1088/2041-8205/807/1/L12}, 807, L12

\bibitem[\protect\citeauthoryear{{Ocvirk} et~al.,}{{Ocvirk}
  et~al.}{2018}]{Ocvirk2018}
{Ocvirk} P.,  et~al., 2018, arXiv e-prints, p. arXiv:1811.11192

\bibitem[\protect\citeauthoryear{{Okamoto}, {Gao}  \& {Theuns}}{{Okamoto}
  et~al.}{2008}]{Okamoto2008}
{Okamoto} T.,  {Gao} L.,   {Theuns} T.,  2008, \mn@doi [\mnras]
  {10.1111/j.1365-2966.2008.13830.x}, 390, 920

\bibitem[\protect\citeauthoryear{{Okamoto}, {Arimoto}, {Yamada}  \&
  {Onodera}}{{Okamoto} et~al.}{2012}]{Okamoto2012}
{Okamoto} S.,  {Arimoto} N.,  {Yamada} Y.,   {Onodera} M.,  2012, \mn@doi
  [\apj] {10.1088/0004-637X/744/2/96}, 744, 96

\bibitem[\protect\citeauthoryear{{Patra}}{{Patra}}{2018}]{Patra2018}
{Patra} N.~N.,  2018, \mn@doi [\mnras] {10.1093/mnras/sty2167}, 480, 4369

\bibitem[\protect\citeauthoryear{{Pawlik}, {Rahmati}, {Schaye}, {Jeon}  \&
  {Dalla Vecchia}}{{Pawlik} et~al.}{2017}]{Pawlik2017}
{Pawlik} A.~H.,  {Rahmati} A.,  {Schaye} J.,  {Jeon} M.,   {Dalla Vecchia} C.,
  2017, \mn@doi [\mnras] {10.1093/mnras/stw2869}, 466, 960

\bibitem[\protect\citeauthoryear{{Peek} et~al.,}{{Peek}
  et~al.}{2011}]{Peek2011}
{Peek} J.~E.~G.,  et~al., 2011, \mn@doi [\apjs] {10.1088/0067-0049/194/2/20},
  194, 20

\bibitem[\protect\citeauthoryear{{Pontzen} \& {Governato}}{{Pontzen} \&
  {Governato}}{2012}]{Pontzen2012}
{Pontzen} A.,  {Governato} F.,  2012, \mn@doi [\mnras]
  {10.1111/j.1365-2966.2012.20571.x}, 421, 3464

\bibitem[\protect\citeauthoryear{{Pontzen} \& {Tremmel}}{{Pontzen} \&
  {Tremmel}}{2018}]{Pontzen2018}
{Pontzen} A.,  {Tremmel} M.,  2018, \mn@doi [\apjs] {10.3847/1538-4365/aac832},
  237, 23

\bibitem[\protect\citeauthoryear{{Pontzen} et~al.,}{{Pontzen}
  et~al.}{2008}]{Pontzen2008}
{Pontzen} A.,  et~al., 2008, \mn@doi [\mnras]
  {10.1111/j.1365-2966.2008.13782.x}, 390, 1349

\bibitem[\protect\citeauthoryear{{Pontzen}, {Ro{\v{s}}kar}, {Stinson}  \&
  {Woods}}{{Pontzen} et~al.}{2013}]{Pontzen2013}
{Pontzen} A.,  {Ro{\v{s}}kar} R.,  {Stinson} G.,   {Woods} R.,  2013, {pynbody:
  N-Body/SPH analysis for python} (\mn@eprint {ascl} {1305.002})

\bibitem[\protect\citeauthoryear{{Pontzen}, {Tremmel}, {Roth}, {Peiris},
  {Saintonge}, {Volonteri}, {Quinn}  \& {Governato}}{{Pontzen}
  et~al.}{2017}]{Pontzen2017}
{Pontzen} A.,  {Tremmel} M.,  {Roth} N.,  {Peiris} H.~V.,  {Saintonge} A.,
  {Volonteri} M.,  {Quinn} T.,   {Governato} F.,  2017, \mn@doi [\mnras]
  {10.1093/mnras/stw2627}, 465, 547

\bibitem[\protect\citeauthoryear{{Puchwein}, {Haardt}, {Haehnelt}  \&
  {Madau}}{{Puchwein} et~al.}{2019}]{Puchwein2019}
{Puchwein} E.,  {Haardt} F.,  {Haehnelt} M.~G.,   {Madau} P.,  2019, \mn@doi
  [\mnras] {10.1093/mnras/stz222}, 485, 47

\bibitem[\protect\citeauthoryear{{Rasera} \& {Teyssier}}{{Rasera} \&
  {Teyssier}}{2006}]{Rasera2006}
{Rasera} Y.,  {Teyssier} R.,  2006, \mn@doi [\aap]
  {10.1051/0004-6361:20053116}, 445, 1

\bibitem[\protect\citeauthoryear{{Read} \& {Erkal}}{{Read} \&
  {Erkal}}{2019}]{Read2019}
{Read} J.~I.,  {Erkal} D.,  2019, \mn@doi [\mnras] {10.1093/mnras/stz1320},
  487, 5799

\bibitem[\protect\citeauthoryear{{Read}, {Iorio}, {Agertz}  \&
  {Fraternali}}{{Read} et~al.}{2017}]{Read2017}
{Read} J.~I.,  {Iorio} G.,  {Agertz} O.,   {Fraternali} F.,  2017, \mn@doi
  [\mnras] {10.1093/mnras/stx147}, 467, 2019

\bibitem[\protect\citeauthoryear{{Revaz} \& {Jablonka}}{{Revaz} \&
  {Jablonka}}{2018}]{Revaz2018}
{Revaz} Y.,  {Jablonka} P.,  2018, \mn@doi [\aap]
  {10.1051/0004-6361/201832669}, 616, A96

\bibitem[\protect\citeauthoryear{{Rey} \& {Pontzen}}{{Rey} \&
  {Pontzen}}{2018}]{Rey2018}
{Rey} M.~P.,  {Pontzen} A.,  2018, \mn@doi [\mnras] {10.1093/mnras/stx2744},
  474, 45

\bibitem[\protect\citeauthoryear{{Rey}, {Pontzen}  \& {Saintonge}}{{Rey}
  et~al.}{2019a}]{Rey2019a}
{Rey} M.~P.,  {Pontzen} A.,   {Saintonge} A.,  2019a, \mn@doi [\mnras]
  {10.1093/mnras/stz552}, 485, 1906

\bibitem[\protect\citeauthoryear{{Rey}, {Pontzen}, {Agertz}, {Orkney}, {Read},
  {Saintonge}  \& {Pedersen}}{{Rey} et~al.}{2019b}]{Rey2019b}
{Rey} M.~P.,  {Pontzen} A.,  {Agertz} O.,  {Orkney} M. D.~A.,  {Read} J.~I.,
  {Saintonge} A.,   {Pedersen} C.,  2019b, \mn@doi [\apjl]
  {10.3847/2041-8213/ab53dd}, 886, L3

\bibitem[\protect\citeauthoryear{{Ricotti}}{{Ricotti}}{2009}]{Ricotti2009}
{Ricotti} M.,  2009, \mn@doi [\mnras] {10.1111/j.1745-3933.2008.00586.x}, 392,
  L45

\bibitem[\protect\citeauthoryear{{Ricotti} \& {Gnedin}}{{Ricotti} \&
  {Gnedin}}{2005}]{Ricotti2005}
{Ricotti} M.,  {Gnedin} N.~Y.,  2005, \mn@doi [\apj] {10.1086/431415}, 629, 259

\bibitem[\protect\citeauthoryear{{Rorai}, {Carswell}, {Haehnelt}, {Becker},
  {Bolton}  \& {Murphy}}{{Rorai} et~al.}{2018}]{Rorai2018}
{Rorai} A.,  {Carswell} R.~F.,  {Haehnelt} M.~G.,  {Becker} G.~D.,  {Bolton}
  J.~S.,   {Murphy} M.~T.,  2018, \mn@doi [\mnras] {10.1093/mnras/stx2862},
  474, 2871

\bibitem[\protect\citeauthoryear{{Rosdahl} \& {Blaizot}}{{Rosdahl} \&
  {Blaizot}}{2012}]{Rosdahl2012}
{Rosdahl} J.,  {Blaizot} J.,  2012, \mn@doi [\mnras]
  {10.1111/j.1365-2966.2012.20883.x}, 423, 344

\bibitem[\protect\citeauthoryear{{Rosdahl}, {Blaizot}, {Aubert}, {Stranex}  \&
  {Teyssier}}{{Rosdahl} et~al.}{2013}]{Rosdahl2013}
{Rosdahl} J.,  {Blaizot} J.,  {Aubert} D.,  {Stranex} T.,   {Teyssier} R.,
  2013, \mn@doi [\mnras] {10.1093/mnras/stt1722}, 436, 2188

\bibitem[\protect\citeauthoryear{{Rosdahl} et~al.,}{{Rosdahl}
  et~al.}{2018}]{Rosdahl2018}
{Rosdahl} J.,  et~al., 2018, \mn@doi [\mnras] {10.1093/mnras/sty1655}, 479, 994

\bibitem[\protect\citeauthoryear{{Roth}, {Pontzen}  \& {Peiris}}{{Roth}
  et~al.}{2016}]{Roth2016}
{Roth} N.,  {Pontzen} A.,   {Peiris} H.~V.,  2016, \mn@doi [\mnras]
  {10.1093/mnras/stv2375}, 455, 974

\bibitem[\protect\citeauthoryear{{Ryan-Weber}, {Begum}, {Oosterloo}, {Pal},
  {Irwin}, {Belokurov}, {Evans}  \& {Zucker}}{{Ryan-Weber}
  et~al.}{2008}]{RyanWeber2008}
{Ryan-Weber} E.~V.,  {Begum} A.,  {Oosterloo} T.,  {Pal} S.,  {Irwin} M.~J.,
  {Belokurov} V.,  {Evans} N.~W.,   {Zucker} D.~B.,  2008, \mn@doi [\mnras]
  {10.1111/j.1365-2966.2007.12734.x}, 384, 535

\bibitem[\protect\citeauthoryear{{Sawala} et~al.,}{{Sawala}
  et~al.}{2016}]{Sawala2016}
{Sawala} T.,  et~al., 2016, \mn@doi [\mnras] {10.1093/mnras/stv2597}, 456, 85

\bibitem[\protect\citeauthoryear{{Shapiro}, {Giroux}  \& {Babul}}{{Shapiro}
  et~al.}{1994}]{Shapiro1994}
{Shapiro} P.~R.,  {Giroux} M.~L.,   {Babul} A.,  1994, \mn@doi [\apj]
  {10.1086/174120}, 427, 25

\bibitem[\protect\citeauthoryear{{Simon}}{{Simon}}{2019}]{Simon2019}
{Simon} J.~D.,  2019, \mn@doi [\araa] {10.1146/annurev-astro-091918-104453},
  57, 375

\bibitem[\protect\citeauthoryear{{Smith}, {Sijacki}  \& {Shen}}{{Smith}
  et~al.}{2019}]{Smith2019}
{Smith} M.~C.,  {Sijacki} D.,   {Shen} S.,  2019, \mn@doi [\mnras]
  {10.1093/mnras/stz599}, 485, 3317

\bibitem[\protect\citeauthoryear{{Somerville}}{{Somerville}}{2002}]{Somerville2002}
{Somerville} R.~S.,  2002, \mn@doi [\apjl] {10.1086/341444}, 572, L23

\bibitem[\protect\citeauthoryear{{Su} et~al.,}{{Su} et~al.}{2018}]{Su2018}
{Su} K.-Y.,  et~al., 2018, \mn@doi [\mnras] {10.1093/mnras/sty1928}, 480, 1666

\bibitem[\protect\citeauthoryear{{Susa} \& {Umemura}}{{Susa} \&
  {Umemura}}{2004}]{Susa2004}
{Susa} H.,  {Umemura} M.,  2004, \mn@doi [\apj] {10.1086/379784}, 600, 1

\bibitem[\protect\citeauthoryear{{Teyssier}}{{Teyssier}}{2002}]{Teyssier2002}
{Teyssier} R.,  2002, \mn@doi [\aap] {10.1051/0004-6361:20011817}, 385, 337

\bibitem[\protect\citeauthoryear{{Theuns}, {Leonard}, {Efstathiou}, {Pearce}
  \& {Thomas}}{{Theuns} et~al.}{1998}]{Theuns1998}
{Theuns} T.,  {Leonard} A.,  {Efstathiou} G.,  {Pearce} F.~R.,   {Thomas}
  P.~A.,  1998, \mn@doi [\mnras] {10.1046/j.1365-8711.1998.02040.x}, 301, 478

\bibitem[\protect\citeauthoryear{{Tollerud} \& {Peek}}{{Tollerud} \&
  {Peek}}{2018}]{Tollerud2018}
{Tollerud} E.~J.,  {Peek} J.~E.~G.,  2018, \mn@doi [\apj]
  {10.3847/1538-4357/aab3e4}, 857, 45

\bibitem[\protect\citeauthoryear{{Toro}, {Spruce}  \& {Speares}}{{Toro}
  et~al.}{1994}]{Toro1994}
{Toro} E.~F.,  {Spruce} M.,   {Speares} W.,  1994, \mn@doi [Shock Waves]
  {10.1007/BF01414629}, 4, 25

\bibitem[\protect\citeauthoryear{{Upton Sanderbeck}, {D'Aloisio}  \&
  {McQuinn}}{{Upton Sanderbeck} et~al.}{2016}]{UptonSanderbeck2016}
{Upton Sanderbeck} P.~R.,  {D'Aloisio} A.,   {McQuinn} M.~J.,  2016, \mn@doi
  [\mnras] {10.1093/mnras/stw1117}, 460, 1885

\bibitem[\protect\citeauthoryear{{Weisz} et~al.,}{{Weisz}
  et~al.}{2012}]{Weisz2012}
{Weisz} D.~R.,  et~al., 2012, \mn@doi [\apj] {10.1088/0004-637X/748/2/88}, 748,
  88

\bibitem[\protect\citeauthoryear{{Weisz}, {Dolphin}, {Skillman}, {Holtzman},
  {Gilbert}, {Dalcanton}  \& {Williams}}{{Weisz} et~al.}{2014}]{Weisz2014}
{Weisz} D.~R.,  {Dolphin} A.~E.,  {Skillman} E.~D.,  {Holtzman} J.,  {Gilbert}
  K.~M.,  {Dalcanton} J.~J.,   {Williams} B.~F.,  2014, \mn@doi [\apj]
  {10.1088/0004-637X/789/2/147}, 789, 147

\bibitem[\protect\citeauthoryear{{Wheeler} et~al.,}{{Wheeler}
  et~al.}{2019}]{Wheeler2019}
{Wheeler} C.,  et~al., 2019, \mn@doi [\mnras] {10.1093/mnras/stz2887}, 490,
  4447

\bibitem[\protect\citeauthoryear{{Wright}, {Brooks}, {Weisz}  \&
  {Christensen}}{{Wright} et~al.}{2019}]{Wright2019}
{Wright} A.~C.,  {Brooks} A.~M.,  {Weisz} D.~R.,   {Christensen} C.~R.,  2019,
  \mn@doi [\mnras] {10.1093/mnras/sty2759}, 482, 1176

\bibitem[\protect\citeauthoryear{{de Jong} et~al.,}{{de Jong}
  et~al.}{2008}]{deJong2008}
{de Jong} J.~T.~A.,  et~al., 2008, \mn@doi [\apj] {10.1086/587835}, 680, 1112

\makeatother
\end{thebibliography}



\appendix
\section{Self-shielding prescription} \label{app:shielding}

In this Appendix, we describe the self-shielding prescription implemented in our `Fiducial' model (see also \citealt{Aubert2010, Rosdahl2012}). Since we do not track the propagation of ionising radiation , we have to approximate, on the fly, the response of optically thick, self-shielded gas to the UV background. 

For each gas cell with hydrogen density $n_{\text{H}}$, we evaluate the normalised cooling and heating rates at a fictional, enhanced density, $n_{\text{H, boosted}}$ with
\begin{equation}
  n_{\text{H, boosted}} =  n_{\text{H}} \ \exp{\big(\frac{n_{\text{H}}}{n_{\text{H, crit}} }\big)} \, ,
\end{equation}
where $n_{\text{H, boosted}}$ cannot exceed the maximum of the cooling table, and $n_{\text{H, crit}}$ defines the threshold density at which gas starts self-shielding. Gas with density close or above $n_{\text{H, crit}}$ is therefore associated an exponentially increased neutral fraction and cooling rate -- and exponentially damped photo-heating and ionization rate -- compared to its optically-thin value, thus mimicking the effect of self-shielding.

The value of $n_{\text{H, crit}}$ is fixed at all times and positions, in this work to $0.01 \, \cmcube$. It therefore neglects the physical dependency of self-shielding on the column density (rather than number density) of neutral hydrogen, as well as local ionizing conditions. Nonetheless, this chosen threshold was shown to provide a good match to the gas neutral fractions obtained with radiative transfer simulations (e.g. \citealt{Pontzen2008, FaucherGiguere2010, Rosdahl2012}).


\bsp	
\label{lastpage}
\end{document}